\documentclass[10pt]{article}

\usepackage[utf8]{inputenc}
\usepackage[T1]{fontenc}
\usepackage{lmodern}
\usepackage{microtype}
\usepackage[margin=2cm]{geometry}
\usepackage{graphicx}
\usepackage{amsmath,amssymb,amsthm}
\usepackage[super,sort&compress]{natbib}
\usepackage[hyphens]{url}
\usepackage{booktabs}
\usepackage{caption}
\usepackage{subcaption}
\usepackage{enumerate}
\usepackage{fancyhdr}
\usepackage{lineno}
\usepackage{xcolor}
\usepackage{authblk}
\usepackage{multicol}

\usepackage{amsmath}
\usepackage{amssymb}
\usepackage{amsthm}
\usepackage{mathabx} 
\usepackage{dsfont} 
\usepackage{xcolor} 
\usepackage{graphicx}
\usepackage{subcaption}
\usepackage{booktabs}
\usepackage{tabularx,ragged2e}
\newcolumntype{C}{>{\Centering\arraybackslash}X} 
\usepackage{authblk} 
\usepackage{appendix}
\usepackage[citecolor=blue,colorlinks=true,linkcolor=blue]{hyperref}

\setcitestyle{citesep={,}}

\makeatletter
\renewcommand{\@maketitle}{%
  \newpage
  \null
  \vskip 1em%
  \begin{center}%
    \let \footnote \thanks
    {\Large\bfseries \@title \par}%
    \vskip 1em%
    {\normalsize
      \lineskip .5em%
      \begin{tabular}[t]{c}%
        \@author
      \end{tabular}\par}%
    \vskip 1em%
    {\normalsize \@date}%
  \end{center}%
  \par
  \vskip 1em}
\makeatother

\usepackage{titlesec}
\titleformat{\section}
  {\normalfont\large\bfseries}{\thesection}{1em}{}
\titleformat{\subsection}
  {\normalfont\normalsize\bfseries}{\thesubsection}{1em}{}

\renewenvironment{abstract}{%
  \small
  \begin{center}
    \bfseries Abstract
  \end{center}
  \quotation
}{%
  \endquotation
}

\newenvironment{keywords}{%
  \small
  \textbf{Keywords:} 
}{%
  \par
}

\begin{document}

\title{Mapping the political landscape from data traces: multidimensional opinions of users, politicians and media outlets on X}

\author[a,b,c,$\dagger$]{Antoine Vendeville}
\author[a,$\dagger$]{Jimena Royo-Letelier}
\author[d,a,c]{Duncan Cassells}
\author[a]{Jean-Philippe Cointet}
\author[a]{Maxime Crépel}
\author[a]{Tim Faverjon}
\author[a]{Théophile Lenoir}
\author[a]{Béatrice Mazoyer}
\author[a]{Benjamin Ooghe-Tabanou}
\author[e,f,a]{Armin Pournaki}
\author[b,g]{Hiroki Yamashita}
\author[b,a,c,*,$\dagger$]{Pedro Ramaciotti}

\affil[a]{Sciences Po médialab, Paris, France}
\affil[b]{Complex Systems Institute of Paris Ile-de-France CNRS, Paris, France}
\affil[c]{Learning Planet Institute, Learning Transitions Unit, CY Cergy Paris University, Paris, France}
\affil[d]{Sorbonne Université, CNRS, LIP6, Paris, France}
\affil[e]{Max Planck Institute for Mathematics in the Sciences, Leipzig, Germany}
\affil[f]{Laboratoire Lattice, École Normale Supérieure - PSL - CNRS - Univ.\ Sorbonne Nouvelle, Montrouge, France}
\affil[g]{École des Hautes Études en Sciences Sociales, Paris, France}

\date{}

\maketitle

\begin{abstract}
Studying political activity on social media often requires defining and measuring political stances of users or content.
Relevant examples include the study of opinion polarization, or the study of political diversity in online content diets.
While many research designs rely on operationalizations best suited for the US setting, few allow addressing more general political systems, in which users and media outlets might exhibit stances on multiple ideology and issue dimensions, going beyond traditional Liberal-Conservative or Left-Right scales.
To advance the study of more general online ecosystems, we present a dataset pertaining to a population of X/Twitter users, parliamentarians, and media outlets embedded in a political space spanned by dimensions measuring attitudes towards immigration, the EU, liberal values, elites and institutions, nationalism and the environment, in addition to left-right and liberal-conservative scales. We include indicators of individual activity and popularity: mean number of posts per day, number of followers, and number of followees.
We provide several benchmarks validating the positions of these entities and discuss several applications for this dataset.
\end{abstract}

\begin{center}
\end{center}


\vspace{1.0cm}

\begin{multicols}{2}

\section{Introduction}
The study of phenomena related to public opinion online has become relevant to many fields, from social and political sciences \citep{mccoy2018polarization,baldassarri2008partisans,reiljan2020fear,boxell2024crosscountry,orhan2022relationship,iyengar2019origins,fiorina2008political,finkel2020political,druckman2021affective,kozlowski2021issue,lelkes2016mass}, to computer science and complex systems \citep{peralta2024multidimensional,bakshy2015exposure,garimella2017longterm,conover2011political,flamino2023political,pham2021balance,baumann2021emergence}. However, because many pioneering studies originated in the US, researchers have traditionally produced more studies and data considering opinions as binary choices (e.g., Democrat- vs Republican-leaning), or as positions along single-dimensional political frames (Liberal-Conservative or Left-Right scales). This framework will not be suited to more general cases, for example countries where more political dimensions are required to analyze national politics \citep{mccoy2018polarization,bakker2012exploring,dolezal2013structure,stoll2010elite,bornschier2010new}.

Most landmark works have relied on large datasets of behavioral traces collected from online platforms or panels of users matched to additional data or surveys. X, in particular, has been extensively studied \citep{ozkula2023easy,kubin2021role}, due to the combination of a large active user base (including most of the prominent political figures in the Western world), a profusion of political content, and an accessible and well-documented API. However, research efforts are now impeded as the API was officially locked behind an expensive paywall in February 2023. Data is not easily accessible either on other major platforms such as Reddit or Facebook, while smaller platforms are often restricted to fringe user bases \citep{mekacher2023systemic,papasavva2021voat,papasavva2020raiders} or do not propose well-documented APIs \citep{corso2024what,mekacher2024koo}. Research efforts can be supported through collaboration with the companies themselves \citep{nyhan2023likeminded,bakshy2015exposure,guess2023howdo,guess2023reshares,gonzalezbailon2023asymmetric}, but this approach raises concerns related to independence and reproducibility \citep{wagner2023independence,roozenbeek2022democratize}. Finally, despite the large quantity of empirical social media research, few researchers release publicly the datasets they collect. 


We address these limitations with the release of a large-scale, high-granularity and anonymized dataset containing multi-dimensional opinions of $N=978,933$ users in the French political sphere on X, including all of the 883 French Members of Parliament who had accounts by February 2023. The political opinions of users in our dataset are provided in the form of real numbers that indicate individual positions on $d=16$ ideological scales and issue dimensions, including the Left-Right axis, anti-elite sentiment, attitudes towards nationalism, immigration, E.U. integration, and environmental policies. Using links pointing towards press articles shared by the users, we also provide positions of 400 popular French websites, including most relevant media outlets, blogs, and forums. Finally, we include individual indicators of activity and popularity for the users, as well as indicators of popularity for the media outlets.

To derive the opinions of users, we applied the methodology proposed by Ramaciotti et al.\ (2022) \citep{ramaciotti2022inferring}. This methodology first performs a multidimensional ideology scaling \citep{clinton2004statistical} on a bipartite follower network linking MPs with their followers \citep{barbera2015birds}. This network was built via a comprehensive collection of follow links in February 2023. Following the method by Ramaciotti et al.\ (2022) \citep{ramaciotti2022inferring}, we then selected a political survey dataset to calibrate and identify the multidimensional latent positions resulting from the ideology scaling procedure. 
For this, we selected the 2019 and 2023 Chapel Hill Expert Surveys (CHES) \citep{ches2019,ches2023} to map the output of the multidimensional ideology scaling onto the dimensions or scales of the CHES datasets. This second step lets us overcome the traditional identification problem of ideology scaling, which embeds users in latent spaces that do not necessarily correspond to actual ideological or political axes. The first dimension of such spaces is often a good proxy of the most salient line of division in national politics, but it can fail to produce readily usable political positions in national settings that display several salient and independent lines of division. 

The methodology we use is solely based on follow relationships, to the exclusion of textual data. To validate the contents of our dataset we look at textual self-descriptions provided by the users on their X profile bios, which are independent from data used for the inference of positions. We annotated these profiles according to the political stances that they express, both with human annotators and with generative AI annotation protocols. These annotations used in validation are also provided in the dataset. The positions of the web domains that we include in the dataset (computed on the bases of the positions of users that shared posts with URL from these domains) are in good agreement with a categorical classification of the most important French media on the Left-Right dimension by previous research \citep{corpora_paper,corpora_data}. The use of political positions of media outlets is well-known to the scientific literature in many fields, and the dataset here presented enables researchers to conduct many of these investigations considering now several relevant ideology and issue dimensions, beyond Left-Right or Liberal-Conservative scales.

Opinion inference based on social media data counts several and diverse methods.
Other relevant publicly available datasets include: a French X dataset collected during the 2017 Presidential Elections, comprising retweet and mention counts between 20K users manually labeled with their preferred political party \citep{fraisier2018elysee,fraisier2018elysee_data}; an American dataset of popular web domains and political leanings of Facebook users sharing them in 2014 \citep{bakshy2015exposure,bakshy2015exposure_data}; a dataset comprising survey responses and web-browsing data collected from voluntary participants in several countries in 2021-2022 \citep{torcal2023dynamics,torcal2023dynamics_data}; a dataset comprising over 200K messages posted in Indian Whatsapp groups during the 2019 General Elections, with 3.8K of these messages manually labeled to indicate support or disapproval of political parties \citep{srivastava2021poliwam,srivastava2021poliwam_data}; a dataset of American political blogs and hyperlinks between one another, collected during the 2004 Presidential Elections \citep{adamic2005political,adamic2005political_data}. 

In comparison with most previously available datasets for studying phenomena related to political opinions in online settings, our main contribution is the release of the first public dataset with multidimensional and continuous opinion estimates, alongside indicators of activity and popularity, for almost a million X users, parliamentarians and hundreds of media domains. 

Next, we present the methodology for the construction of the dataset, a detailed description of the data records and files included in the dataset, validation benchmarks for the ideological and issue positions provided by the dataset, and, finally, a brief discussion of limitations and the data and code availability statements.

\section{Methodology}

Abundant evidence suggests that the choice of which political actors to follow is an informative signal of ideology \citep{barbera2015birds,ramaciotti2021unfolding,ramaciotti2022inferring,ramaciotti2020your,bond2015quantifying,barbera2015understanding,briatte2015recovering}. The methodology we use for opinion inference relies on the follow relationships that exist between Members of Parliament (MPs) and the general public on X, relying on the existence of political homophily on the part of users choosing to follow MPs. We start by describing the data collection process. Then, we detail the opinion inference methodology, which unfolds in two steps. Next, we explain how we derived individual indicators of activity and popularity. Finally, we explain how we derived positions for a selection of French media domains on ideology and issue dimensions.

\subsection{Collection of the MP-followers network}

The data collection process reproduces that done by Ramaciotti et al., (2022), in which they used data collected in 2019.
In February 2023, we identified the X profiles of all 886 French MPs present on the platform (out of 925 in the French parliament), and collected all their followers using the software \texttt{minet 1.00.0-a15} \citep{plique2023minet}. Both chambers of the parliament are concerned, the lower (\textit{Assemblée nationale}) and the upper (\textit{Sénat}) houses. We include MPs who resigned from their mandate before December 2022, as well as their substitutes\footnote{MPs who resigned include mostly those who accepted a position in the government after the elections. Resigned MPs and their substitutes amount to $N=46$ of the MPs present in the data.}. Their followers were then filtered based on the criteria from Barber\'a \citep{barbera2015birds}, keeping only those who followed at least 3 MPs. This ensures that the users left in our dataset have a sufficient knowledge of national politics. Individuals that are knowledgeable about national politics are more accurately represented in their political opinion by spatial models, a property related to \textit{political sophistication} \citep{luskin1990explaining}. 

Barber\'a \citep{barbera2015birds} also suggests removing users with less than 25 followers, to filter out bots and inactive accounts. To allow for fine-grained studies of the relationships between popularity and opinions we keep these accounts in the data, and propose an alternative dataset with those users excluded (see data availability). We show in the Appendix that keeping these accounts did not alter significantly the computed political positions.

We remove MPs who are left without followers after this filtering step. We end up with $M=883$ MPs and $N=978,206$ followers. From there, we build a directed bipartite MP-follower network, yielding a binary adjacency matrix $A\in\{0,1\}^{N\times M}$. Importantly, follower-follower and MP-MP connections are ignored. The network contains 9,601,175 edges, MPs have average in-degree 10,910 and followers have average out-degree 10.

Similar studies performing latent ideological inference often rely on retweet networks \cite{falkenberg2022growing,falkenberg2024patterns}. While retweet links may also provide a source of homophilic choice data, they could also lead to biased results due to agenda setting and the dynamics of issue saliency over time. The database of tweets that we rely on for the computation of media positions, covers in particular the period of the invasion of Ukraine by Russia, and the widely contested pension reform in France. Such significant political event have the capability to strongly restructure the political discussion online, resulting in a biased representation of the overall political space. Using follow links on the other hand, ensures that we are capturing a stable political space which we expect to underpin the general dynamics of the public debate in France. Additionally, our collection of follow links is exhaustive, and therefore does not suffer from potential issues due to data sampling. These issues could arise when retrieving individual tweets and retweets from the now defunct Twitter API \citep{morstatter2021sample,tromble2021where,dibona2024sampled}.

\subsection{Inference of latent ideological positions}

As Ramaciotti et al., (2022) describe in their method, we first apply the ideology scaling method originally proposed by Barber{\'a} \citep{barbera2015birds} for extending the use of ideology scaling to social media data. It takes as input a bipartite network of connections between users and MPs, and outputs positions in a latent homophily space in which MPs that are close in space are followed by similar sets of users, and, conversely, in which users that are close in space follow a similar set of MPs.
The method assumes that the observation (i.e., the choices of whom to follow) obeys a probabilistic homophily law in which followers decide to follow MPs on the bases of unobservable positions on some latent space:
\begin{equation}
    \begin{split}
    \text{Prob}(A_{ij}=1\vert \alpha_i,\beta_j,\gamma,\phi_i,\phi_j) = \\\text{logit}^{-1}(\alpha_i+\beta_j-\gamma\| \phi_i - \phi_j\|^2), 
    \end{split}
    \label{eq:homophily}
\end{equation}
where $A_{ij}=1$ when user $i$ follows user $j$, $\alpha_i$ and $\beta_j$ quantify respectively the tendency to follow others and the tendency to be followed of users $i$ and $j$, and $\gamma$ is a shape parameter that controls the relative weight between homophilic ($\phi_i$ and $\phi_j$) and idiosyncratic parameters ($\alpha_i$ and $\beta_j$). The network of connections is known, and the goal is to infer the unknown latent ideological positions $\phi$ that best explain this network. These positions will then reflect ideological similarities and differences between the users. 

Using a Markov Chain Monte Carlo (MCMC) estimation procedure on the network of U.S. parliamentarians and their X followers, Barberà \citep{barbera2015birds} showed that the latent positions $\phi$ stand as good indicators of ideological positions. 
It has been later shown that the inference of the parameters of the equation can be approximated via a Correspondence Analysis (CA) \citep{greenacre2017correspondence}. CA is a dimensionality reduction procedure for categorical data, which has been shown to be a fast and efficient approximation method for the computation of $\phi$ \citep{barbera2015birds,lowe2008understanding,caroll1997equivalence}. 

We perform a CA on the adjacency matrix $A$ to obtain individual ideological positions in a latent space.
The dimensionality of this latent space can be fixed between 1 and 883.
Because we will map positions from this latent space to the space subtended by the dimensions of the political survey we use for calibration---the Chapel Hill Expert Survey data (CHES), see below---and because this map is computed with the positions of political parties on both spaces, the choice of the number of dimensions for the latent space depends on the number of available political parties on both spaces.
Our 883 MPs embedded in the latent space can be identified with 11 parties from the CHES data. 
We compute proxy multidimensional positions of each party in the latent space as the average position of MPs from that party.
We remark that positions along dimensions of this latent space may be readily linked to positions on ideologies and issues, but this cannot be assured because of the lack of identification for values $\phi$ in the probabilistic equation (in particular with respect to isometries such as translations and rotations).

\subsection{Mapping latent positions onto ideology and issue dimensions from political surveys}
\label{subsec:mapping_latent_positions}

Following the method described by Ramaciotti et al. (2022), we map these latent multidimensional positions onto the ideology and issue dimensions of a political survey instrument used as reference.
We use two political survey instruments administered around the time of our data collection: the 2019 \citep{ches2019} and the 2023 \citep{ches2023} CHES data.
In these surveys, political scientists estimate the positions of the main political parties of multiple European countries on multiple ideology and issue dimensions (e.g.\ placement on the Left-Right scale, or attitudes towards immigration policy), on a continuous scale from $0$ to $10$\footnote{Dimensions pertaining to E.U.\ integration scale from 1 to 7, but we rescale them to $[0-10]$ for the sake of consistency.}. 
Crucially, positions on these dimensions are endowed with spatial reference frames.
For instance, on the Left-Right dimension, 0 stands for the leftmost position for political parties, 10 for the rightmost, and 5 is the political center.
While CHES 2019 is older and less consistent with the date of collection of our data, it contains a much larger set of dimensions (51 versus 11 for CHES 2023), and both waves include 4 dimensions in common (Left-Right, GAL-TAN, EU, and Anti-elite), which we also use as validation (see Fig.~\ref{fig:2019_vs_2023}).
Our approach does not seek to infer positions for our dataset on all available CHES dimensions, and we focus instead on those for which we can propose validation tests (see Table~\ref{tab:dimensions}).

\begin{table*}
\centering
\scriptsize
\caption{List of dimensions from the Chapel Hill Expert Survey (CHES) that we consider in the method to calibrate the political positions of X users, with statistics summarizing these positions. The reference points for dimensions are between values 0 and 10 (the extreme positions for political parties in the CHES data), while individuals can be positioned outside these bounds (outliers).}
\begin{tabular}{|llcccr|}
\hline
\textbf{CHES dimension} & \textbf{Description} & \textbf{CHES wave} & \textbf{Mean} & \textbf{Std.} & \textbf{\% outliers} \\
\hline
lrgen & Left - Right & 2019 & 6.308 & 2.293 & 3.078 \\
corrupt\_salience & Importance of reducing political corruption & 2019 & 4.708 & 0.691 & 0.000 \\
people\_vs\_elite & Opposes - Favors (direct democracy) & 2019 & 4.802 & 0.995 & 0.023 \\
immigrate\_policy & Favors - Opposes (immigration) & 2019 & 6.861 & 1.819 & 4.008 \\
sociallifestyle & Favors - Opposes (liberal policies) & 2019 & 5.494 & 2.185 & 0.592 \\
nationalism & Cosmopolitanism - Nationalism & 2019 & 6.355 & 2.030 & 5.518 \\
antielite\_salience & Anti-elite sentiment & 2023 & 6.771 & 1.875 & 2.081 \\
eu\_position & Anti EU - Pro EU & 2023 & 4.832 & 1.978 & 0.923 \\
lrecon & Left - Right (economy) & 2023 & 5.445 & 1.320 & 0.033 \\
refugees & Opposes - Favors (welcoming Ukrainian refugees) & 2023 & 5.822 & 1.820 & 0.407 \\
galtan & Liberal - Conservative & 2023 & 6.004 & 1.902 & 0.209 \\
environment & Favors - Opposes (environment over economy) & 2019 & 5.645 & 1.073 & 0.190 \\
lrecon & Left - Right (economy) & 2019 & 5.350 & 1.958 & 0.049 \\
antielite\_salience & Anti-elite sentiment & 2019 & 7.040 & 1.880 & 3.320 \\
eu\_position & Anti EU - Pro EU & 2019 & 4.705 & 2.275 & 0.540 \\
galtan & Liberal - Conservative & 2019 & 5.988 & 1.584 & 0.026 \\
\hline
\end{tabular}
\label{tab:dimensions}
\end{table*}

As described in the work detailing the positioning method \citep{ramaciotti2022inferring}, we compute an affine transformation mapping between our latent space and the space of political dimensions spanned by CHES.
This transformation is fitted by using positions of political parties in both the latent space and the survey dimensions.
Party positions come readily available in the CHES datasets.
Following Ramaciotti et al. (2022), we compute party positions in the latent space as the mean position of MPs from each party.
For each CHES dimension of interest $d_c$, we fit an affine transformation between the multidimensional positions of political parties in the latent space onto the party positions along $d_c$. We use a Ridge regression \citep{hoerl1970ridge}, with penalty parameter $\alpha=1.0$.
We restrain the number dimensions of the departure space to $P-1$, with $P$ being the number of political parties that exist both in the latent space and in the survey dataset.
This value, $P-1$, is the number of dimensions that would fully determine the affine transformation under no regularization \citep{ramaciotti2022inferring}.
From the parties of the MPs that were manually annotated, 11 were present in the CHES 2023 data, (resp.\ 9 in CHES 2019), leading to fitting affine transformations using the first 10 and 8 dimensions of the latent space, respectively.
See the Appendix and Table~\ref{tab:parties} for a more detailed description of the parties.

Once the affine transformations are fitted this way, we apply them to the latent positions of each follower and MP. This way, we obtain coordinates for all followers and MPs in the political space spanned by the chosen CHES dimensions. Due to the regularization we apply, the positions of the political parties in our data---computed as the average position of the relevant MPs---may differ slightly from their positions in CHES. This deviation is small in the sense that the average Pearson correlation between party position computed with the affine transformation and given by the CHES data is  0.94. The mean absolute difference in party positions is 0.63 (for reference, CHES dimensions span from 0 to 10).

\subsection{Activity and popularity} 

We include in this dataset metrics of popularity and activity for users positioned in our multidimensional ideology and issue space. These metrics are obtained or calculated from the user metadata obtained during the collection process using the platform's API. They comprise the mean number of posts per day, the number of followers, and the number of friends (also known as followees or followed accounts), at collection date. We compute for each user the mean number of posts per day, dividing their total number of posts by the number of days elapsed between account creation and our date of collection.

\subsection{Media sharing collection} 

To compute positions of media outlets, we use a database of tweets collected during 18 months between 1 January 2022 and 30 June 2023 via the search query \texttt{lang:fr filter:links}. The database contains all tweets published in French during that period which contained at least one URL. When a tweet was part of a thread we also collected the entire thread, including the original post. For the collection, we used the Gazouilloire tool version 1.2.0 \citep{ooghe2023gazouilloire} running query calls to X's search API v1.1. We aggregate this way an average of 2.6 million tweets each day and a total of 1.4 billion tweets over the whole period\footnote{Data collected live was lost for the months of January and February 2023 and had therefore to be recollected a few months later, resulting in a little less tweets per day over those two months due to the known attrition of posts either deleted by X or their authors themselves.}.

We filter this database of tweets and keep only original posts produced by the users for which we estimated a political position in Section~\ref{subsec:mapping_latent_positions}, and containing URLs towards a domain belonging to the French media ecosystem.
This design choice achieves two objectives.
Filtering out quotes, we capture how individuals with a given issue and ideological position propose URL sources, without this source being leveraged or elicited by the context of a discussion unfolding on a thread.
Filtering out retweet or shares, we capture behavior in which individuals have personally curated (and thus read with higher probability), instead of effortlessly sharing a post that may have contained a URL source (which they may have not read). 
We define this ecosystem as a collection of 747 domains, taken from ref.\ \citep{defacto_data} and corresponding to major French media sources---newspapers, regional press, radio and TV channels, blogs and other digital information outlets. The methodology for selection of the domains is detailed in ref.\ \citep{defacto_paper}. 
Filtering the collection of tweets this way, we obtain a collection of 3,429,848 tweets by 71,374 followers. There were 747 domains identified in the original study, of which 692 we find in at least one tweet from our users. To guarantee a good signal quality, we discard media domains whose articles were not tweeted by at least 100 different users. We end up with 400 media domains from which 3,272,574 URLs were shared by 70,604 followers (7.2\% of all followers collected before). For each media domain, we provide the number of posts containing an URL from the domain, and the number of users in our collection who emitted these posts.  
Under this operationalization, the affinity between author and cited source may be of recognition of authority and relevance (i.e., the author deems the source worthy of mention, even if it is to critice the source or the content), ideology and issue affinity, or a combination of both. In the Validation section, we disentangle these possibilities by measuring the issue and ideology regularity of the authors of tweets including URLs from a given source. We also show that the media positions we derive are strongly correlated with the results of an independent study positioning French media outlets on the Left-Right axis.

\subsection{Computing positions of media domains on survey dimensions}

For each media domain, and for each one of the selected issue and ideology dimensions from the CHES data, we compute the position of a media in a dimension as the mean position of the followers that cited URLs from the domain.
We do not take into account the number of tweets per user but solely a binary variable that takes value 1 if the user ever tweeted a link to a webpage from the domain, and 0 otherwise.
This design choice is motivated by the fact that polarized users tend to be more active on social media. 
We seek to capture who finds a source worthy of mention, and not how many times.
Operationalized this way, the computed positions must be interpreted as those of the public that deems the corresponding media worthy of citing.
These positions do not necessarily represent the positions of contributors, journalists and editors of the media, which can display considerable heterogeneity.
These data on media positions are useful as a measure of attention and authority granted by individuals.
For studies that would want to leverage these positions for studies related to media bias, we also compute and provide in the data standard deviations and perform Hartigan's dip test \citep{hartigan1985dip} to evaluate the spread and the potential bimodality of the opinion distributions across users who share each domain. Using these additional metrics, researchers can select medias that are highly cited by users with coherent positions in our selected issue and ideology dimensions.

\begin{figure*}[!t]
    \centering
    \begin{subfigure}[b]{0.49\textwidth}
         \centering
         \includegraphics[width=\textwidth]{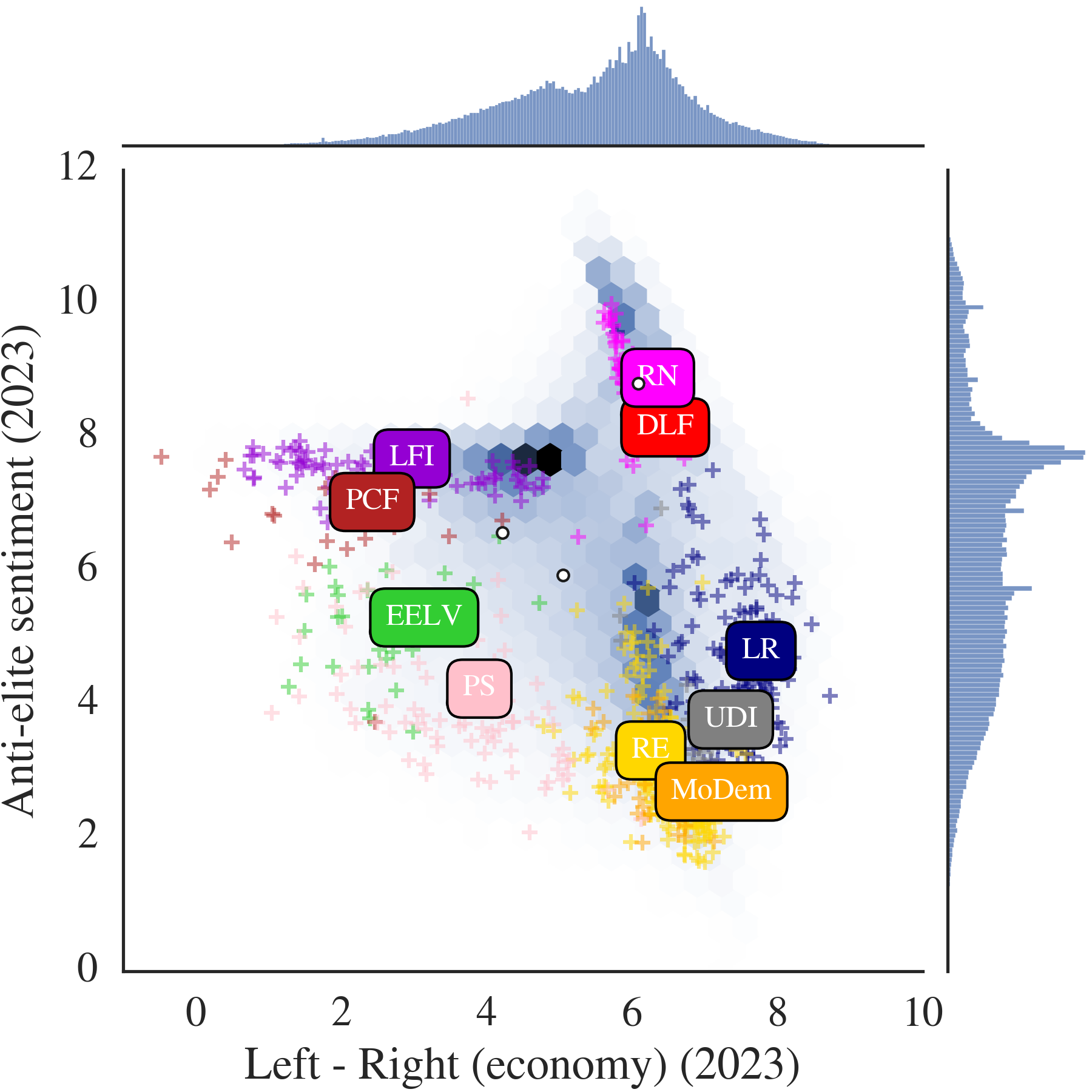}
     \end{subfigure}~
     \begin{subfigure}[b]{0.49\textwidth}
         \centering
         \includegraphics[width=\textwidth]{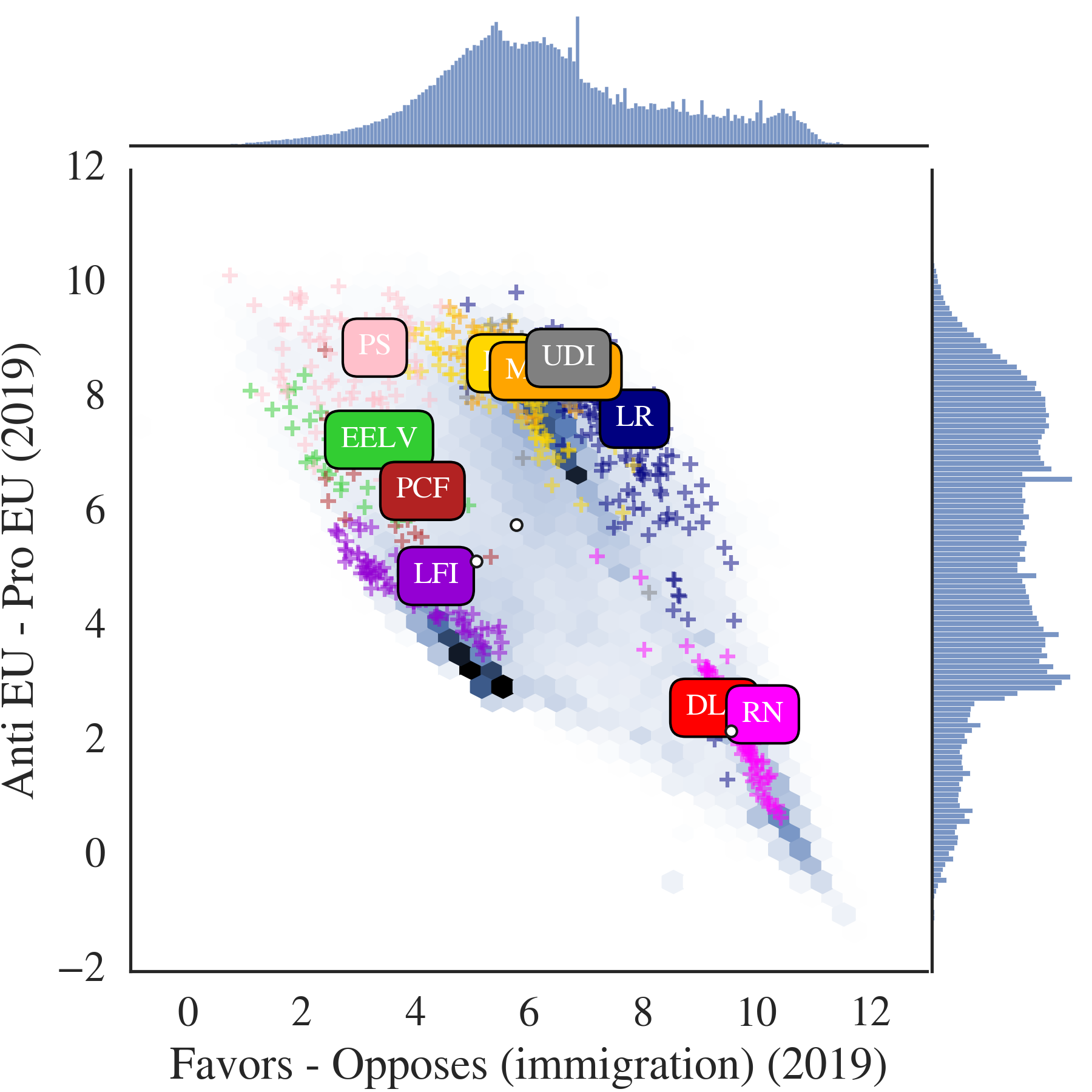}
     \end{subfigure}~
    \caption{Positioning of users, MPs, parties and three selected media domains on four political axes: economic Left-Right and anti-elite sentiment (left), European integration and ecology (right). Colored crosses represent MPs. Colored rectangles indicate the average position of each party's parliamentarians. Marginal densities and hexagons indicate the distribution of followers positions. White dots indicate the positions of three selected media domains.}
    \label{fig:bidimensional}
\end{figure*}

Finally, to give an example illustration of the data we release, Fig.~\ref{fig:bidimensional} shows the positions of MPs, parties and three major media domains along four selected dimensions.

\section{Data records}

The released files contain seven tables in CSV format, one Jupyter notebook and several images. The CSV files are the following:
\begin{itemize}
    \item \texttt{mps\_positions.csv}
    \item \texttt{followers\_positions.csv}
    \item \texttt{followers\_human\_annotations.csv}
    \item \texttt{followers\_llm\_annotations.csv}
    \item \texttt{mps\_activity.csv}
    \item \texttt{followers\_activity.csv}
    \item \texttt{domains\_positions.csv}
\end{itemize}
We detail below the size and content of these tables. Most of them exhibit sets of columns pertaining to similar features declined across the multiple political dimensions and annotation labels present in our data ; we do not list all such columns in the tables below but give a representative example and the total number of such columns each time. All floats are given with three decimals, except for mean\_tweets\_per\_day (five decimals). Names of MPs are included, but other users have been anonymized and all information allowing for potential identification has been removed. For the sake of practicality, we endowed each user and MP with a unique random \emph{pseudo id} that allows for matching data across the different tables.

\subsection{Positions of MPs} 

The file \texttt{mps\_positions.csv} contains the names, party affiliations and political positions of MPs. It has 883 rows and 19 columns. Each row corresponds to one MP that had a platform account at the time of our manual annotation procedure. The columns are detailed in Table~\ref{tab:mp_csv}.

\begin{table*}[!ht]
\setlength\extrarowheight{2pt} 
\centering
\caption{Detail of columns for \texttt{mps\_positions.csv}.}
\begin{tabularx}{\textwidth}{|lXr|}
\hline
\textbf{Column name} & \textbf{Description} & \textbf{Format} \\ 
\hline
pseudo\_id & Unique (artificial) identifier of the user. & string \\ \hline
name & First name and surname of the MP. & string \\ \hline
party & Party affiliation of the MP. & string \\ \hline
xxx\_yy (e.g.\ lrecon\_23) & Position on dimension xxx in the space of CHES 20yy. 16 such columns. & float \\
\hline
\end{tabularx}
\label{tab:mp_csv}
\end{table*}

\subsection{Positions of followers of MPs} 

The file \texttt{followers\_positions.csv} contains the political positions of the followers. It has 978,050 rows and 17 columns. The columns are the same as for \texttt{mps\_positions.csv} minus the party and name columns. Each row corresponds to one user.

\subsection{Human annotations of followers bios} 

The file \texttt{followers\_human\_annotations.csv} contains human annotations of the bios of followers, used for the validation. It has 5,864 rows and 9 columns---users with all NaN annotations were discarded. Each row corresponds to one follower. These annotations correspond to labels produced by a set of human annotators, prompted to read profile bios of users in our dataset, and annotate whether they display information allowing us to classify them into groups relevant for our validation metrics. The columns are detailed in Table~\ref{tab:follower_human_annot}.

\begin{table*}[!ht]
\setlength\extrarowheight{2pt} 
\centering
\caption{Detail of columns for \texttt{followers\_human\_annotations.csv}.}
\begin{tabularx}{\textwidth}{|lXX|}
\hline
\textbf{Column name} & \textbf{Description} & \textbf{Format} \\ \hline
pseudo\_id & Unique (artificial) identifier of the user. & string \\ 
\hline
xxx (e.g.\ eurosceptic) & Human annotations of the bios for class xxx. 8 such columns. & 1.0 (identified as xxx), nan (unidentified). \\ 
\hline
\end{tabularx}
\label{tab:follower_human_annot}
\end{table*}

\subsection{LLM annotations of profile bios of followers} 

The file \texttt{followers\_llm\_annotations.csv} contains annotations of the followers bios produced by LLMs and used for the validation.
It has 317,684 rows and 16 columns---users with all NaN annotations were discarded. Each row corresponds to one follower. The columns are detailed in Table~\ref{tab:follower_llm_annot}.

\begin{table*}[!ht]
\setlength\extrarowheight{2pt} 
\centering
\caption{Detail of columns for \texttt{followers\_llm\_annotations.csv}.}
\begin{tabularx}{\textwidth}{|lXX|}
\hline
\textbf{Column name} & \textbf{Description} & \textbf{Format} \\
\hline
pseudo\_id & Unique (artificial) identifier of the user. & string \\ \hline
xxx & LLM annotations of the bios for class xxx. 15 such columns. & 0.0 (identified as not xxx), 1.0 (identified as xxx), nan (unidentified). \\
\hline
\end{tabularx}
\label{tab:follower_llm_annot}
\end{table*}

\subsection{Activity and popularity of MPs} 

The file \texttt{mps\_activity.csv} contains the names, party affiliations, activity-related and popularity-related metrics for the MPs. It has 883 rows and 6 columns. Each row corresponds to one MP. All MPs are included. The columns are detailed in Table~\ref{tab:mp_activity}.

\begin{table*}[!ht]
\setlength\extrarowheight{2pt} 
\centering
\caption{Detail of columns for \texttt{mps\_activity.csv}.}
\begin{tabularx}{\textwidth}{|lXr|}
\hline
\textbf{Column name} & \textbf{Description} & \textbf{Format} \\ 
\hline
pseudo\_id & Unique (artificial) identifier of the user. & string \\ \hline
name & First name and surname of the MP. & string \\ \hline
party & Party affiliation of the MP. & string \\ \hline
mean\_tweets\_per\_day & Average number of posts per day. & float \\ \hline
followers & Number of followers. & float \\ \hline
followees & Number of followees. & float \\
\hline
\end{tabularx}
\label{tab:mp_activity}
\end{table*}

\subsection{Activity and popularity of followers} 

The file \texttt{followers\_activity.csv} contains activity and popularity metrics for the followers. It has 978,050 rows and 4 columns. Each row corresponds to one follower. Some followers did not appear in our database of tweets and are excluded from the table. The columns are detailed in Table~\ref{tab:follower_activity}.

\begin{table*}[!ht]
\setlength\extrarowheight{2pt} 
\centering
\caption{Detail of columns for \texttt{followers\_activity.csv}.}
\begin{tabularx}{\textwidth}{|lXr|}
\hline
\textbf{Column name} & \textbf{Description} & \textbf{Format} \\ 
\hline
pseudo\_id & Unique (artificial) identifier of the user. & string \\ \hline
mean\_tweets\_per\_day & Average number of posts per day. & float \\ \hline
followers & Number of followers. & float \\ \hline
followees & Number of followees. & float \\
\hline
\end{tabularx}
\label{tab:follower_activity}
\end{table*}

\subsection{Positions of media domains} 

The file \texttt{domains\_positions.csv} contains the positions of the media domains. It has 400 rows and 84 columns. Each row corresponds to one media domain. The rows are sorted in order of decreasing user\_count (see below). The columns are detailed in Table~\ref{tab:domains_csv}.

\begin{table*}[!ht]
\setlength\extrarowheight{2pt} 
\scriptsize
\caption{Detail of columns for \texttt{domains.csv}.}
\begin{tabularx}{\textwidth}{|lXr|}
\hline
\textbf{Column Name} & \textbf{Description} & \textbf{Format} \\ 
\hline
domain & Web address of the domain. & string \\ \hline
media\_category & Media category from ref.~\citep{corpora_data}. & string \\ \hline
user\_count & Number of users who tweeted URLs pointing towards this domain. & integer \\ \hline
tweet\_count & Number of tweets containing URLs pointing towards this domain. & integer \\ \hline
xxx\_yy\_mean (e.g.\ lrecon\_23\_mean) & Average position of users who shared URLs from this domain on dimension xxx in the space of CHES 20yy. 16 such columns. & float \\ \hline
xxx\_yy\_std (e.g.\ lrecon\_23\_std) & Standard deviation of the positions of users who shared URLs from this domain. 16 such columns. & float \\ \hline
xxx\_yy\_quantile (e.g.\ lrecon\_23\_quantile) & Quantile (20\%) calculated from the distribution of the values of columns xxx\_yy\_mean across all domains. 16 such columns. & integer \\ \hline
xxx\_yy\_dip (e.g.\ lrecon\_23\_dip) & Statistic of Hartigan's dip test \citep{hartigan1985dip} performed on the distribution of the positions of users who shared URLs pointing toward this domain, in the corresponding dimension. 16 such columns. & float \\ \hline
xxx\_yy\_pval (e.g. lrecon\_23\_pval) & p-value of Hartigan's dip test. 16 such columns. & float \\ 
\hline
\end{tabularx}
\label{tab:domains_csv}
\end{table*}

\section{Validations}

In this section we present several computations showing the quality and consistency of the positions computed for users and for medias.
Our method for position estimation is purely structural, in that it relies solely on the \emph{follow} relationships that exist between MPs and other users on X, and on relations between medias and users whenever they include them in their tweets.
Notably, our methods so far exclude textual data.
To demonstrate the robustness of our results, we compare the positions we computed with political stances stated by the users through text in their profile bios (text self-descriptions curated by users to appear at the top of their personal profiles in the platform).
In these bios, users may often convey, as we show below, information accusing their ideological stances (e.g., presenting themselves explicitly as Left- or Right-leaning individuals) and their stances on issues (e.g., explicitly taking position on immigration or environmental protection).

We now present the protocols through which we identified users expressing political stances in their profile bios, and how we leveraged them in computing metrics showing the quality of the ideology and issue positions in our dataset.
All of the results and figures presented in this section can be reproduced using the code provided with the data. The exact text of bios are not provided in the data to protect the anonymity of users.

\subsection{Annotation of profile bios}

For each of the selected ideology and issue dimensions, we establish a pair of labels associated with a distinguishable leaning or stances of users, for which the dimension should provide an order.
For instance, using the text profile bios, we label each user as being Left-leaning (``left'') or Right-leaning (``right'') whenever possible (admittedly, not all users hint at their leaning on their profiles) to quantify the degree to which the continuous Left-Right dimension we inferred orders these users. For the E.U.\ dimension, we annotate users as``eurosceptic'' or ``pro-european'' whenever possible. Dimensions and the corresponding annotation labels are listed in Table~\ref{tab:validation}.
We aim to identify users that have distinguishable self-reported leanings and stances, and not all possible ways in which users convey their stances.
In other words, our labels are suited for our validation protocol, but do not exhaust the many possible ways in which users hint at their political positions through text in their bios.
This strategy borrows from that of previous ideology scaling using follower networks by Barbera \cite{barbera2015birds}, in which he identified users self-identifying as liberal or conservative (among other labels) to validate their estimated ideological positions.
Our strategy expands that of Barbera, adapting a greater number of dimensions, similar to what Ramaciotti et al. \cite{ramaciotti2024attitudinally} proposed for validating two dimensions: Left-Right and anti-elite positions.
We compute these labels in two independent ways: using LLM automatic annotation, and, for fewer dimensions (because of the involved cost), using human annotations.
See in Table~\ref{tab:validation} the labels used and the corresponding dimensions in which we used them for validation.

The criteria and instructions used for annotation (both for automatic and human) rely on the description of dimensions contained in the political survey data we use for calibration, i.e., the CHES data. The CHES data codebooks contain detailed descriptions of the substance of each dimension. 
Some dimensions do not span between dychotomous extremes (like Left-Right), but measure the degree to which a given attitude is held.
The most notable example is the CHES anti-elite dimension, measuring the degree of adoption of anti-elite and anti-establishment populist rhetoric.
In this case, to produce dichotomous labels, we label users as being from ``elite'' groups or displaying ``populist'' rhetoric to validate the Anti-elite dimension. Individuals displaying populist rhetoric (e.g., subscribing to a vision of society divided into the ``people'' and ``elites'') may be \textit{also} part of different elites. The validation we propose relies on the fact that \textit{most} users engaging in acute anti-elite rhetoric will not be part of the elite groups, with high probability.

\paragraph{LLM annotation.}
To submit the profile bios to an LLM for annotation, we first translated each follower profile bio text to English using \texttt{M2M100 1.2B} \citep{fan2020englishcentric}, to minimize the probability of the quality of our annotations depending on the language. We then submitted each English-translated bio to the Large Language Model \texttt{zephyr-7B-$\beta$} \citep{tunstall2023zephyr}, producing binary annotations for the collection of political labels considered. The accuracy and utility of LLM annotations for the classification of political content has been examined in several recent studies \citep{heseltine2024large,mens2024positioning}. 
The prompts used for annotation were chosen after an iterative series of tries, manually evaluating the quality of the annotations for a small subset.
Here are two example prompts that we use; an exhaustive list is provided in Appendix.
\begin{itemize}
    \item LLM label ``Left'': \textit{You are an expert in French politics. Please classify the following X profile bio as “Left- leaning” or “Not-Left” according to whether the author of the text (who is from France) is politically Left-leaning or not. The response should be in the form of a single term with the name of the category: “Left-leaning” or “Not-Left”: [TEXT OF THE BIO].}
    \item LLM label ``Eurosceptic'': \textit{You are an expert in European politics. Please classify the following X profile bio as ``Eurosceptic'' or ``Not-Eurosceptic'' according to whether the author of the text (who is from France) holds negative views of the European Union or not. The response should be in the form of a single term with the name of the category: ``Eurosceptic'' or ``Not-Eurosceptic'': [TEXT OF THE BIO]}.
\end{itemize}

Our prompts produce a very large proportion of outputs in the intended requested form. We discarded outputs that do not correspond to one of the two requested allowed categories specified in the prompt. Then we examine LLM annotations produced for pairs of dichotomous labels (``Left'' and ``Right'', ``Pro-European'' and ``Eurosceptic'') to further discard annotations that are contradictory (e.g. users annotated as being both ``Left''- and ``Right''-leaning by the LLM). 
Upon manual examination, we noted that this happened mainly because of ambiguous political messages, e.g.\ ``I am a left-winger because I believe there should not be private property but a right-winger because I believe we should expel immigrants'' (fictional example). Table~\ref{tab:validation_full} reports the number of bios for which we were able to obtain a label.
We compare the annotations for each pair of opposite categories (e.g.\ ``left'' and ``right'', ``populist'' and ``elite'') with the corresponding CHES dimensions (e.g.\ lrecon\_23, antielite\_salience\_23). The full list of correspondence between pairs of labels and CHES dimensions is summarized in Table~\ref{tab:validation}. 

We leveraged annotated profiles in two forms of validation: examination of the monotonicity of the concentration of users with labels along dimensions, and measurement of the order and separation between labeled users along dimensions.

\paragraph{Human annotation.} 
Rather than evaluating LLM annotations with human annotators, we seek to produce an independent measure of validity of estimated positions of users using human annotations.
Because of the cost involved in producing human annotations, we employed a different strategy and protocol.
First, we only considered a few dimensions, namely, Left-Right dimensions (ideological and economic), attitudes towards the EU, immigration, and anti-elite and anti-establishment sentiments.
Second, because cost of human annotations were prohibitive, we adopted neither a purely \textit{descriptive} nor purely \textit{prescriptive} paradigm \cite{rottger2021two}, but a mixed approach.
Instead of providing annotators with the same instructions given to LLMs, we allowed for annotators to propose criteria to define the labels in a two-phase protocol.
The two steps of this descriptive-prescriptive annotation strategy are: 1) first, annotators examine the bios to specify a set of criteria, to then 2) apply these criteria to produce the labels.
While the annotators know the prompts given to LLMs, and thus the number of labels that need to be produced and the main concept underlying them, they are free to propose the rest of the annotation protocol to follow for classifying texts into labels.
Concretely, their proposed protocols allow for them to filter and select the bios that they will then annotate (e.g., based on the presence of keywords they think are important).
This strategy can thus assure only a low rate of false positives and a high rate of true positives, but says little about true and false negatives. More details about the human annotation protocol are provided in the Appendix.
Sections~\ref{subsec:concentration}~\&{}~\ref{subsec:order_separation} demonstrate the convergence validity between LLM and human annotations, and show that both strategies lead to a validation of the estimated positions of users along all targeted CHES dimensions, as quantified by classification metrics.

\subsection{Concentration of labeled users along dimensions}
\label{subsec:concentration}

First, we assess whether labeled users are coherently positioned along the corresponding CHES dimensions. We divide the range of values of each CHES dimension into one-sized bins (0-1, 1-2, 2-3, 3-4, 4-5, 5-6, 6-7, 7-8, 8-9, and 9-10) and group users accordingly. We then proceed to examine the fraction of labeled users among the bins, for the labels corresponding to the selected dimension (Table~\ref{tab:validation}). We also compute Clopper-Pearson confidence intervals \citep{clopper1934theuse} for the fraction of labeled users on each bin.
Our first validation consists in evaluating that the concentration of labeled users is monotonic, and that it increases or decreases coherently across the corresponding dimension.
The figures for 15 labels (each associated to a CHES dimension), are included with the reproducibility material.
The result for a couple of labels is shown here for illustration.

\begin{figure*}[!t]
    \centering
    \begin{subfigure}[b]{0.49\textwidth}
         \centering
         \includegraphics[width=\textwidth]{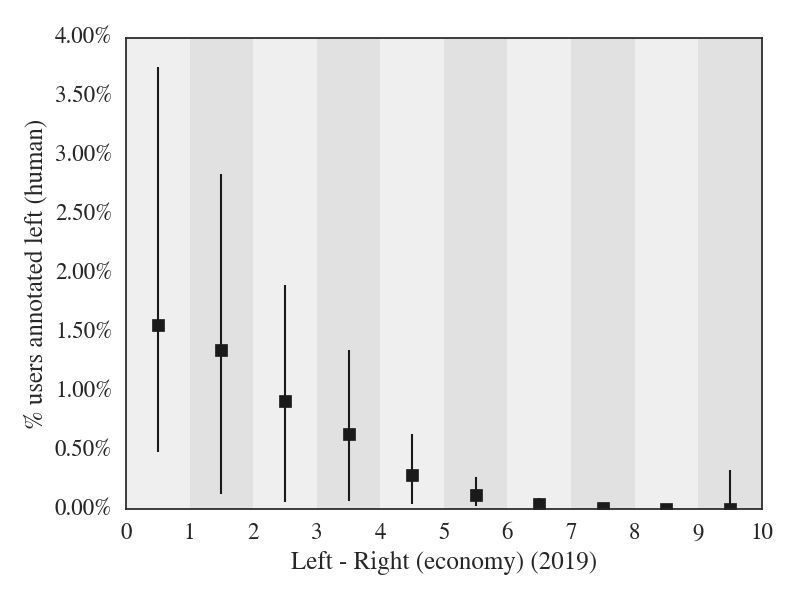}
     \end{subfigure}~
     \begin{subfigure}[b]{0.49\textwidth}
         \centering
         \includegraphics[width=\textwidth]{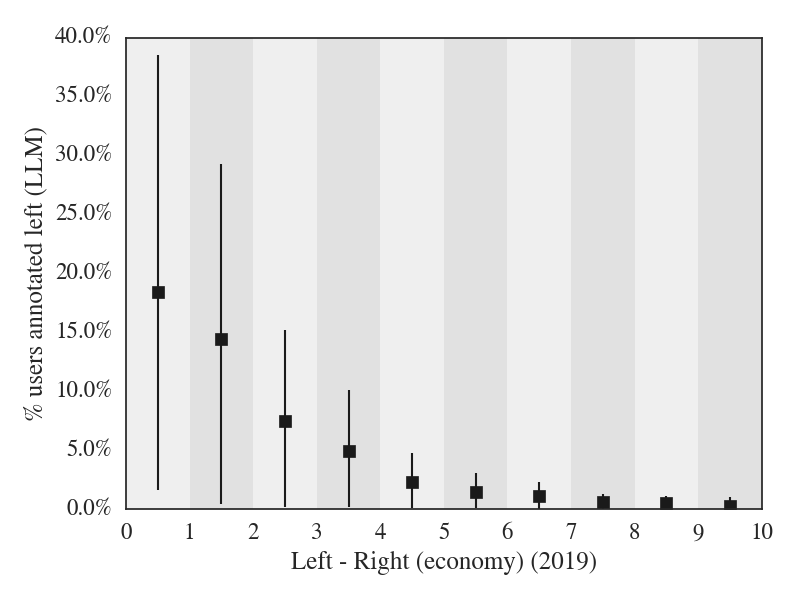}
     \end{subfigure}\\
     \begin{subfigure}[b]{0.49\textwidth}
         \centering
         \includegraphics[width=\textwidth]{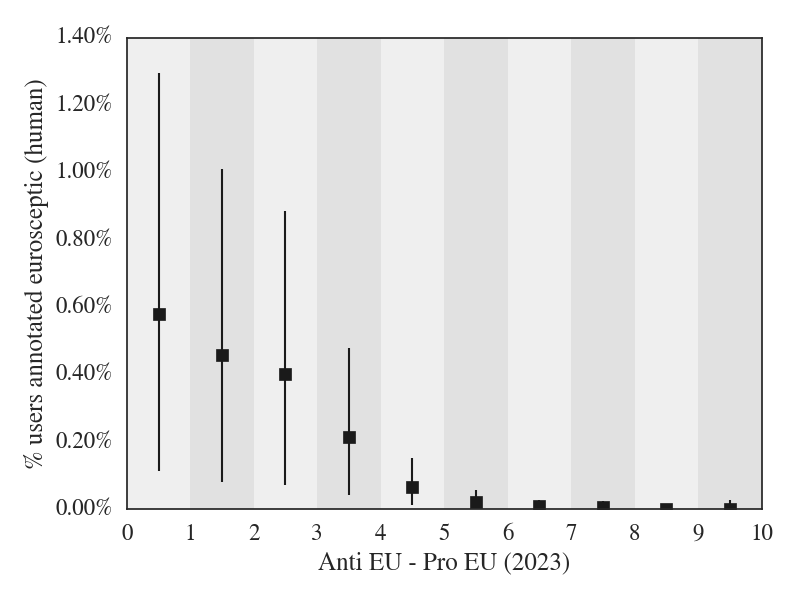}
     \end{subfigure}~
     \begin{subfigure}[b]{0.49\textwidth}
         \centering
         \includegraphics[width=\textwidth]{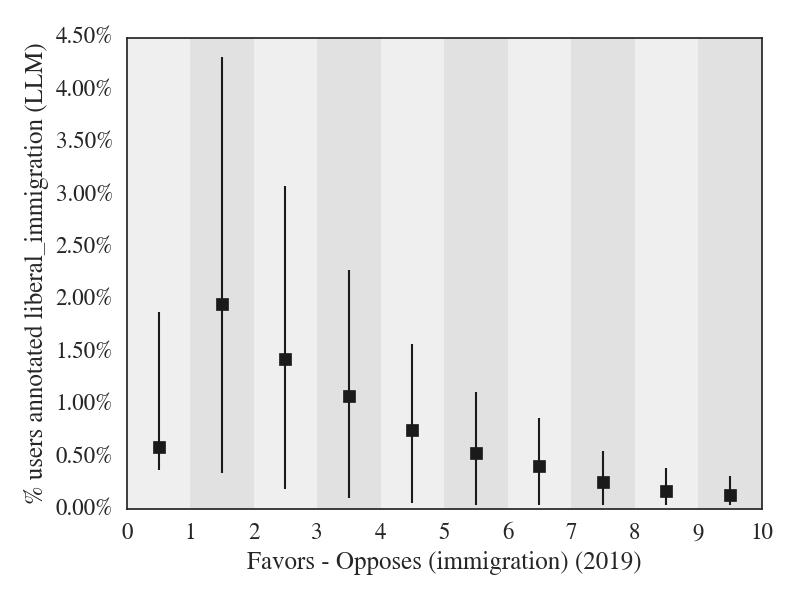}
     \end{subfigure}
    \caption{Proportion of users having a given label indicating ideological or issue stances, as provided by both human and LLM annotations, along the corresponding CHES dimensions they intend to validate. Vertical bars delimit Clopper-Pearson confidence intervals \citep{clopper1934theuse} at the $\alpha=0.05$.}
    \label{fig:bin_plot}
\end{figure*}

Fig.~\ref{fig:bin_plot} shows the distribution of users labeled as having ``left'' leaning (both by human and LLM annotators), showing that their concentration increases monotonically towards lowers values of the CHES Left-Right dimension.
Fig.~\ref{fig:bin_plot} also shows additional examples, including the distribution of users labeled (by humans) as ``eurosceptic'', monotonically concentrated towards lower values of the CHES EU dimension, and users labeled as displaying ``liberal immigration'' stances (by the LLM) along the CHES dimension measuring stances towards immigration.
These selected examples illustrate how human and LLM annotation of text profile bios produce labels for users that are concentrated coherently with the ideology or stances on issues that the CHES dimensions capture.
We find convincing results for all dimensions.
Plots for all combinations of dimension and labels are provided with the data, and can also be reproduced using the reproducibility code available with this article. 

\subsection{Order and separation of labeled users along CHES dimensions}
\label{subsec:order_separation}

We now seek to evaluate how well users with opposite labels are separated along the corresponding CHES dimensions. To this end, we fit a logistic regression model between labels and positions in our opinion space. For the example of ``left'' and ``right'' labeled users along CHES Left-Right dimensions, we consider one label as \textit{failures} and the other as \textit{successes}, we use positions along Left-Right to fit the model, and validate the induced separation and order of labeled users with goodness of fit metrics. 
Because logistic regression is sensible to class imbalance, and because some of our labels are imbalanced, we systematically weight the samples of labeled users by the size of their label class\footnote{We use for this the implementation provided with the \texttt{scikit-learn} library, providing the parameter \texttt{class\_weight='balanced'}.}.
Once we have fitted a logistic regression model, we assess its goodness of fit by measuring ROC AUC and F1 scores \citep{ramaciotti2022measuring}, presented in Table~\ref{tab:validation}. Because F1-score is asymmetric (i.e., achieves different results depending on which label is taken as success), we compute it both ways (i.e., considering one class as success and then as failure) and average the result. Asymmetric F1-scores for each label ordering are provided in Appendix (Table~\ref{tab:validation_full}), while Precision and Recall are computed in the reproducibility code accompanying the data.
In other words, we use the training accuracy of a logistic classifier for pairs of labels on the corresponding dimensions as a goodness of fit metric.

\begin{figure*}[!t]
    \centering
    \begin{subfigure}[b]{0.49\textwidth}
         \centering
         \includegraphics[width=\textwidth]{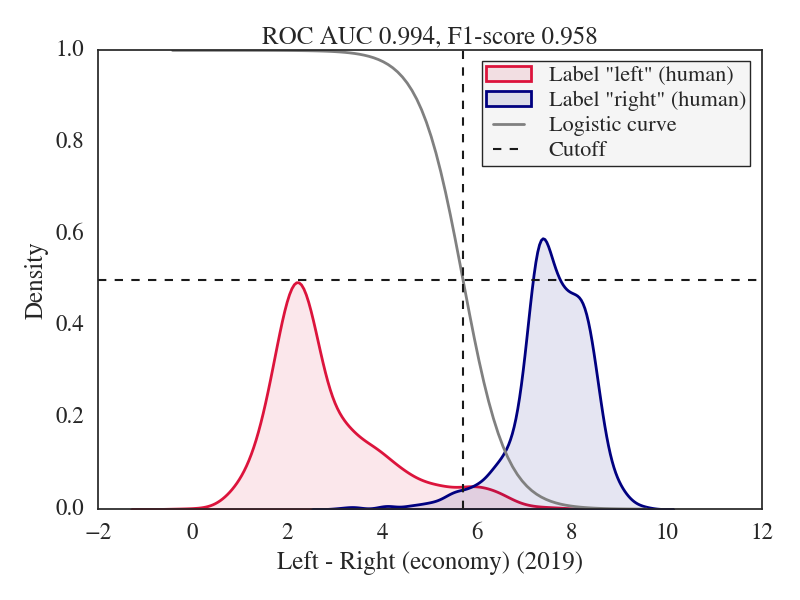}
     \end{subfigure}~
     \begin{subfigure}[b]{0.49\textwidth}
         \centering
         \includegraphics[width=\textwidth]{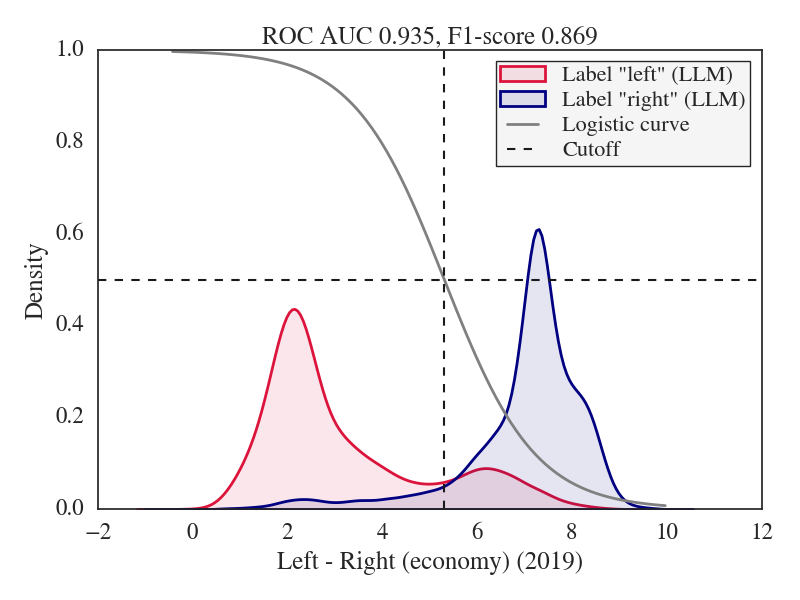}
     \end{subfigure}\\
     \begin{subfigure}[b]{0.49\textwidth}
         \centering
         \includegraphics[width=\textwidth]{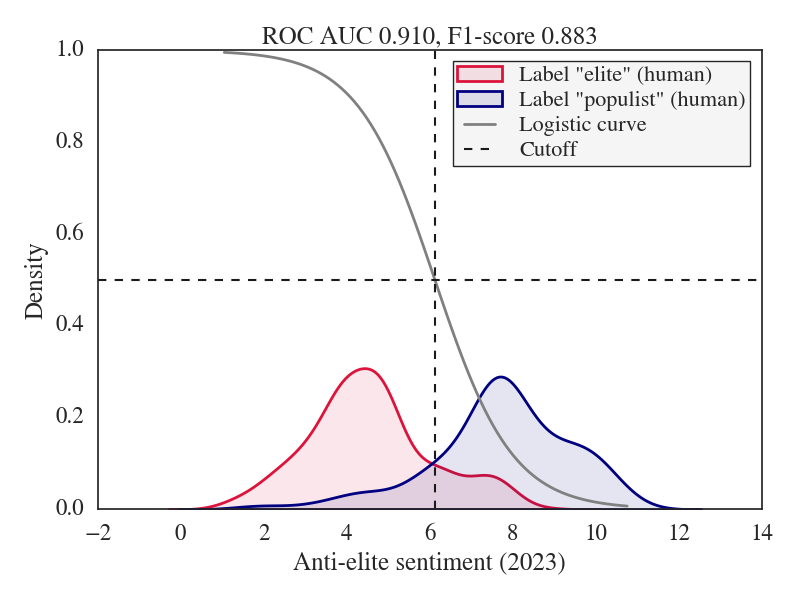}
     \end{subfigure}~
     \begin{subfigure}[b]{0.49\textwidth}
         \centering
         \includegraphics[width=\textwidth]{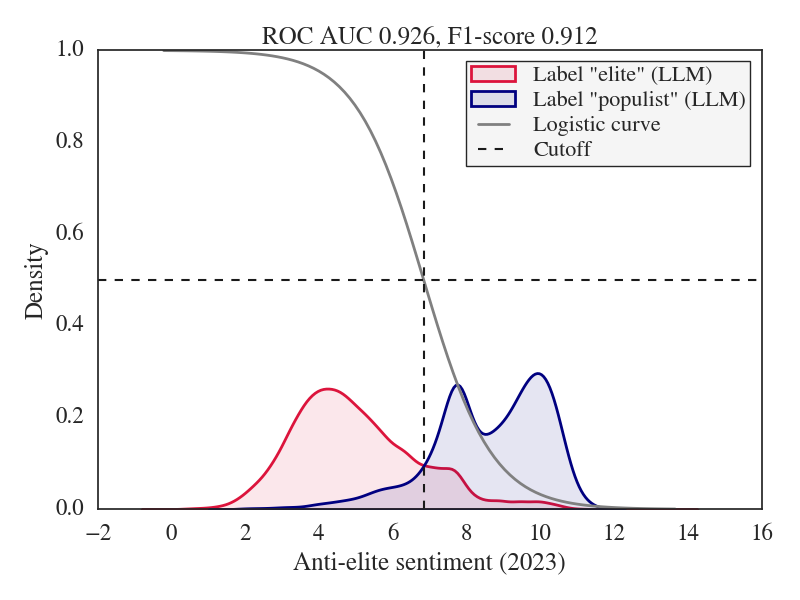}
     \end{subfigure}
    \caption{Illustration of the logistic regression model for the economical left-right dimension (CHES 2019), and the anti-elite sentiment dimension (CHES 2023). We show results obtained with human as well as with LLM annotations. Blue and red areas indicate the distributions of the political position of the annotated users. The cutoff determines where the model changes its prediction. We also indicate ROC AUC and F1 scores.}
    \label{fig:logreg}
\end{figure*}

Upon evaluation, we observe high ROC AUC scores (greater than 0.7) for all the selected dimensions that we test.
For most dimensions we obtained remarkably high scores, greater than 0.9.
Goodness of fit metrics are coherent when comparing human and LLM annotation, although human annotations always achieve better goodness of fit scores.
All of our validations except three achieve F1-score values above 0.6.
Fig.~\ref{fig:logreg} illustrates the goodness of fit metrics on selected examples.

\begin{table*}
\scriptsize
\centering
\caption{Summary of results for goodness of fit for the logistic regression models, sorted by decreasing ROC AUC, serving as main validation of the position of users along selected Chapel Hill Expert Survey (CHES) data dimensions from the 2019 and 2023 waves. Each CHES dimension is validated using users labeled with pairs of dichotomous labels that should be well ordered and separated by the dimension. For the sake of space, ``imm.'' stands for ``immigration''.}
\begin{tabularx}{\textwidth}{|lllXXrr|}
\hline
\textbf{Dimension} & \textbf{Year} & \textbf{Annotator} & \textbf{Label A} & \textbf{Label B} & \textbf{ROC AUC} & \textbf{Avg. F1} \\
\hline
lrgen & 2019 & human & left & right & 0.992 & 0.961 \\
lrecon & 2019 & human & left & right & 0.992 & 0.952 \\
lrecon & 2023 & human & left & right & 0.979 & 0.946 \\
immigrate\_policy & 2019 & human & liberal\_immigration & restrictive\_immigration & 0.972 & 0.924 \\
eu\_position & 2023 & human & eurosceptic & pro\_european & 0.969 & 0.924 \\
eu\_position & 2019 & human & eurosceptic & pro\_european & 0.967 & 0.921 \\
refugees & 2023 & human & liberal\_immigration & restrictive\_immigration & 0.962 & 0.924 \\
galtan & 2023 & LLM & conservative & liberal & 0.943 & 0.882 \\
galtan & 2019 & LLM & conservative & liberal & 0.941 & 0.883 \\
sociallifestyle & 2019 & LLM & conservative & liberal & 0.941 & 0.877 \\
lrgen & 2019 & LLM & left & right & 0.931 & 0.863 \\
lrecon & 2019 & LLM & left & right & 0.929 & 0.859 \\
eu\_position & 2023 & LLM & eurosceptic & pro\_european & 0.927 & 0.823 \\
immigrate\_policy & 2019 & LLM & liberal\_immigration & restrictive\_immigration & 0.922 & 0.864 \\
eu\_position & 2019 & LLM & eurosceptic & pro\_european & 0.917 & 0.803 \\
nationalism & 2019 & LLM & cosmopolitan & nationalist & 0.915 & 0.864 \\
antielite\_salience & 2023 & LLM & elite & populist & 0.915 & 0.697 \\
refugees & 2023 & LLM & liberal\_immigration & restrictive\_immigration & 0.911 & 0.843 \\
antielite\_salience & 2023 & human & elite & populist & 0.903 & 0.824 \\
lrecon & 2023 & LLM & left & right & 0.894 & 0.854 \\
antielite\_salience & 2019 & LLM & elite & populist & 0.886 & 0.684 \\
antielite\_salience & 2019 & human & elite & populist & 0.866 & 0.813 \\
corrupt\_salience & 2019 & LLM & elite & populist & 0.848 & 0.609 \\
people\_vs\_elite & 2019 & LLM & elite & populist & 0.772 & 0.563 \\
corrupt\_salience & 2019 & human & elite & populist & 0.762 & 0.640 \\
people\_vs\_elite & 2019 & human & elite & populist & 0.723 & 0.647 \\
environment & 2019 & LLM & climate\_denialist & pro\_environment & 0.653 & 0.362 \\
environment & 2019 & LLM & economic\_focus & pro\_environment & 0.639 & 0.546 \\
\hline
\end{tabularx}
\label{tab:validation}
\end{table*}

\subsection{Consistency between dimensions present in the CHES 2019 and 2023 waves}

\begin{figure*}[!t]
    \centering
    \includegraphics[width=.75\linewidth]{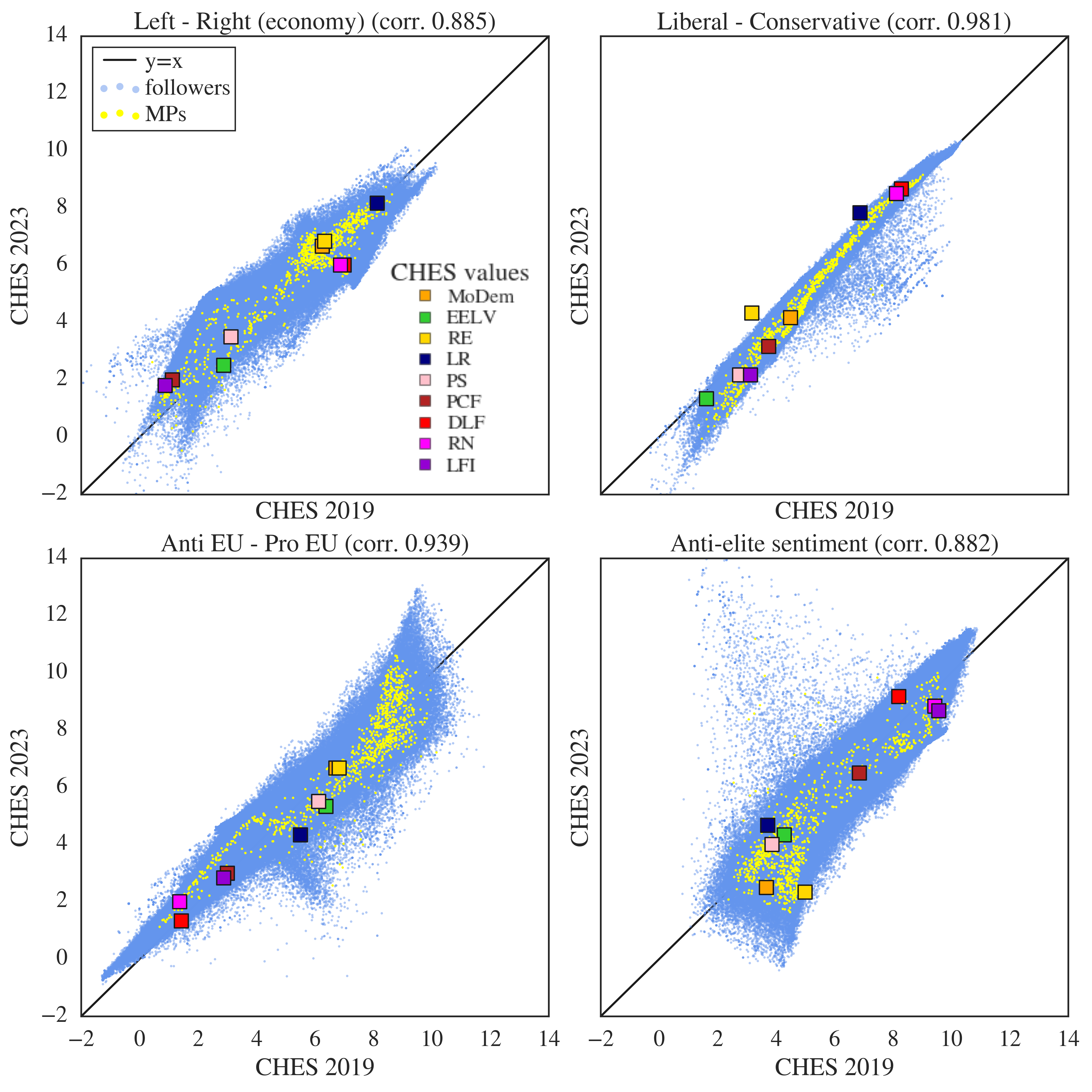}
    \caption{Positions of followers, MPs and party centroids along the four dimensions present in the 2019 and 2023 waves of CHES surveys. Blue dots are followers, yellow dots are MPs, colored squares represent party positions in the CHES surveys. We show the $y=x$ line in black. We also indicate Pearson correlations between the positions of users on the two axes.}
    \label{fig:2019_vs_2023}
\end{figure*}

Four dimensions exist both in the 2019 and 2023 waves of the CHES data.
The 2019 and 2023 are both relevant to our estimates because we used follower links collected in 2023, but produced by users before that.
These dimensions are economic Left-Right, attitudes towards the EU, Liberal-Conservative, and Anti-elite sentiments.
To assess both the reliability of the position inference method we use, and the sensibility of the method to data from these waves, we compare positions of parties, MPs, and their followers.
To compare between positions on a 2019 and a 2023 dimension, we compute the value of Pearson correlation.
As shown in Fig.~\ref{fig:2019_vs_2023}, we find high correlations, indicating consistency across the two surveys (all equal or greater than 0.869).

\subsection{Positions of media outlets}

\begin{figure*}[!t]
    \centering
    \begin{subfigure}[b]{0.49\textwidth}
         \centering
         \includegraphics[width=\textwidth]{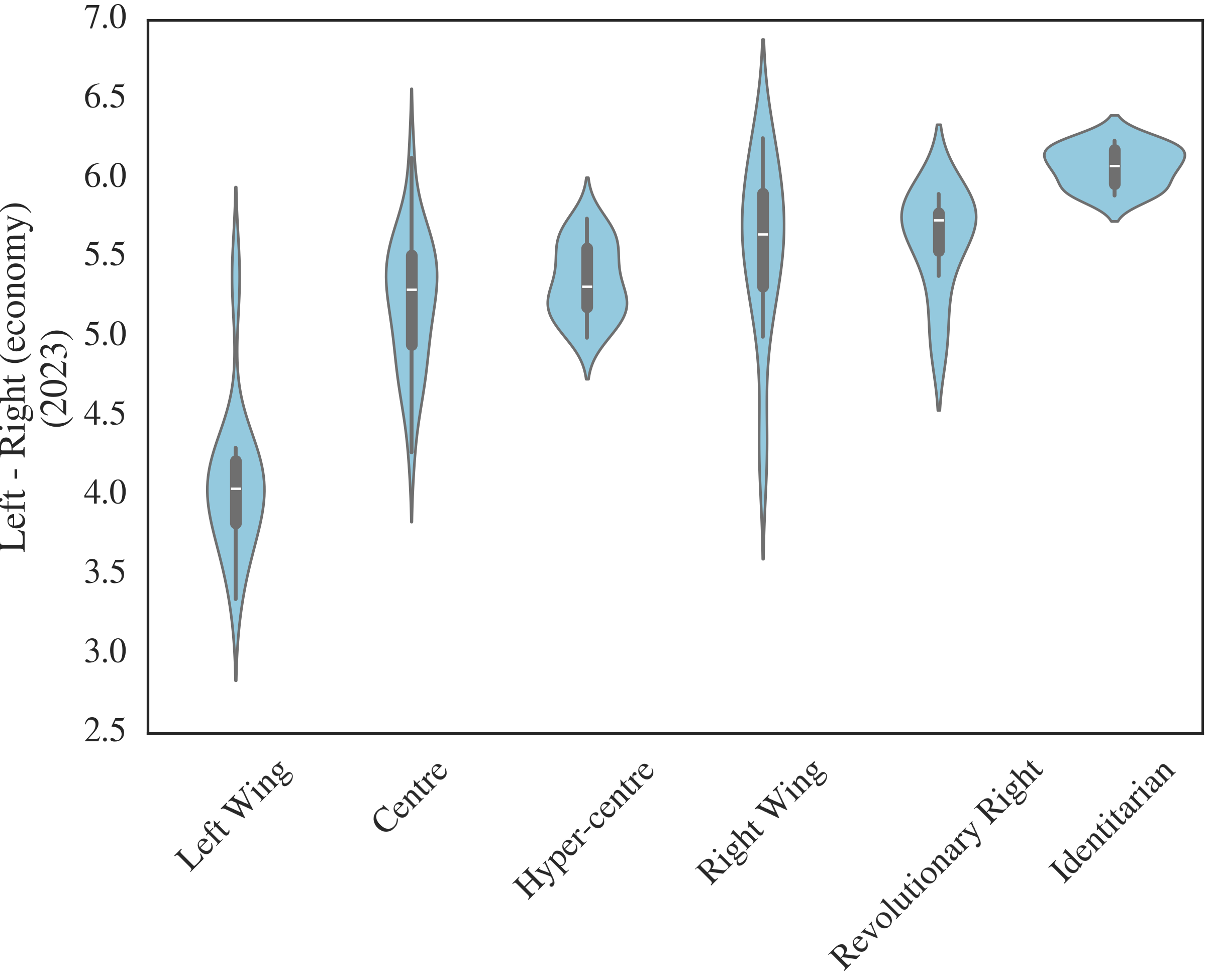}
     \end{subfigure}~
     \begin{subfigure}[b]{0.49\textwidth}
         \centering
         \includegraphics[width=\textwidth]{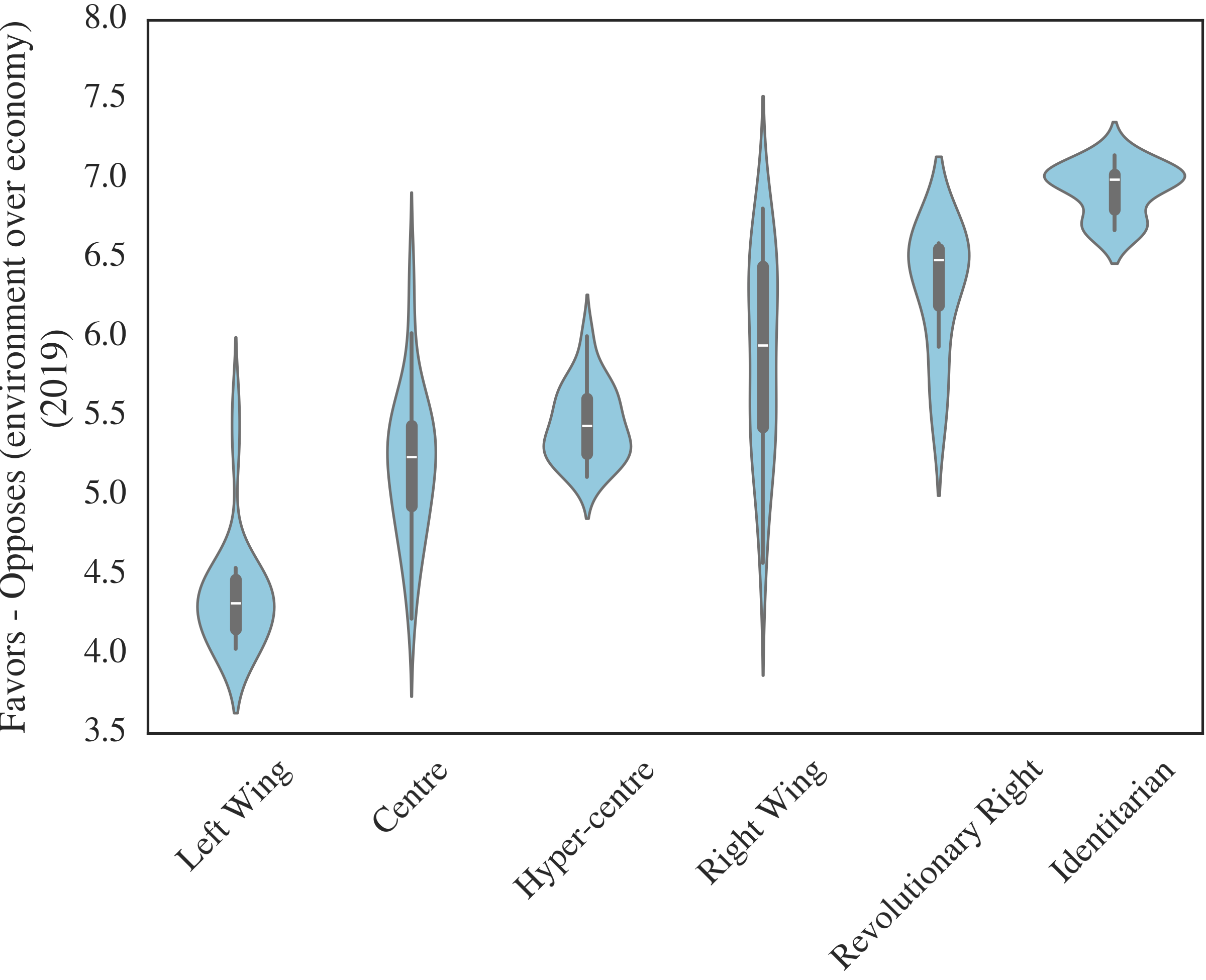}
     \end{subfigure}
    \caption{Validation of the media positions along two political axes. We show the distribution of positions among each media category. These figures demonstrate a strong alignment between the categorization of media along the left-right axis and the economical Left-Right dimension as well as the Environment dimension. The latter is consistent with the difference in environmental discourse between the left and the right in France.}
    \label{fig:media_validation}
\end{figure*}

We now validate the position of the web domains that correspond to French media outlets in our political CHES dimensions. To this end, we leverage a previously curated categorical classification of the most relevant French media outlets by Cointet et al. \citep{corpora_data,corpora_paper}.
This dataset comprises a list of 478 French media domains that users on Twitter (now X) cite most frequently in posts. In this data by Cointet et al., news outlets and politically engaged medias are categorized into the following categories: ``Centre'', ``Hyper-centre'', ``Left Wing'', ``Right Wing'', ``Revolutionary Right'' and ``Identitarian''.
These categories correspond to clusters of the hyperlink citation network of articles from these outlets, computed using the crawled hyperlink network and Stochastic Block Model inference (see the related publication \citep{corpora_paper} for details).
The categories ``Centre'' and ``Hyper-centre'' are not named after the political stance of the corresponding medias, but rather because of the central role they occupy in citation networks, gathering the most important medias that reach a national audience and may thus be read by a large and diverse crowd. 
These central positions are, as shown in Fig.~\ref{fig:media_validation} and other related plots available with the data, tightly related to centrist ideological stances.

Among our 400 media domains, 217 are categorized in the work of Cointet et al. We use the categorization of those domains to validate our political positioning of the domains. For each dimension of the political space, we compare the distribution of the domains positions in the political space across the different categories. We illustrate the result by plotting the distribution of positions along the considered dimension for each category of media. Two examples are shown in Fig.~\ref{fig:media_validation}; additional plots are provided with the data and can also be reproduced with the reproducibility code. 

\section{Limitations and perspectives}

Our dataset offers the potential for several applications. Researchers interested in using this data should however keep in mind that it concerns a specific population, that of X users who follow French Members of Parliament. As online users are in general not representative of the overall population of a country \citep{Mellon2017,Zagheni2015}, conclusions drawn from  analyses of this database should not be generalized to the whole ensemble of French citizens without a thorough evaluation of the validity of such procedure. In fact, this database can also serve as the basis of a proper analysis of the ideological similarities and differences between the online and the general populations, by performing comparisons with respondent surveys such as the \emph{European Social Survey} \citep{ess2023}. 

Because we rely on the Chapel Hill Expert Surveys to position users along the political dimensions, these positions are inherently constrained by the structure of the opinion space subtended by the French political parties.

In the future, this database could be further enriched by subsequent collections to study the evolution of the political space. Further research shall also strive to develop databases of individual political positions for other countries and other online platforms relevant for political communication, such as YouTube, TikTok or Bluesky.

\section{Data availability}

The dataset and the code needed to reproduce analyses in this article are available at \url{https://doi.org/10.17605/OSF.IO/AT5Q2}.

A filtered dataset where users with less than 25 followers are discarded (see Methods) is available at \url{https://doi.org/10.17605/OSF.IO/V28KH}.

\section*{Funding}

This work has been partially funded by the ``European Polarisation Observatory'' (EPO) of CIVICA Research co-funded by EU's Horizon 2020 programme (Grant Agreement No. 101017201), by the project SoMe4Dem funded by EU's Horizon Europe programme (Grant No. 101094752), by the Project Liberty Institute's project ``AI-Political Machines'', and by the \textit{Very Large Research Infrastructure} (TGIR) Huma-Num of CNRS, Aix-Marseille Université and Campus Condorcet.

\section*{CRediT Author Statement}

\textbf{Antoine Vendeville:} Software, Validation, Formal analysis, Investigation, Data curation, Writing, Visualization. \textbf{Jimena Royo-Letelier:} Software, Formal analysis, Resources, Data curation. \textbf{Duncan Cassells:} Data curation. \textbf{Jean-Philippe Cointet:} Data curation. \textbf{Maxime Crépel:} Data curation. \textbf{Tim Faverjon:} Data curation. \textbf{Théophile Lenoir:} Data curation. \textbf{Béatrice Mazoyer:} Resources, Data curation. \textbf{Benjamin Ooghe-Tabanou:} Resources, Data curation. \textbf{Armin Pournaki:} Resources, Data curation. \textbf{Hiroki Yamashita:} Data curation. \textbf{Pedro Ramaciotti:} Conceptualization, Methodology, Investigation, Resources, Data curation, Writing, Visualization, Supervision, Project administration, Funding acquisition.

\section*{Declaration of competing interests}
The authors declare that they have no competing financial interests or personal relationships that could have appeared to influence the work reported in this paper.

\bibliography{references}

@software{plique2023minet,
  author = {Guillaume Plique and Pauline Breteau and Jules Farjas and Héloïse Théro and Jean Descamps and Amélie Pellé and Laura Miguel},
  title = {Minet, a webmining {CLI} tool \& library for {Python}},
  month = apr,
  year = 2023,
  publisher = {Zenodo},
  version = {1.0.0-a15},
  doi = {10.5281/zenodo.7791759},
  url = {https://doi.org/10.5281/zenodo.7791759},
}

@article{boxell2024crosscountry,
  title = {Cross-Country Trends in Affective Polarization},
  author = {Boxell, Levi and Gentzkow, Matthew and Shapiro, Jesse M.},
  year = {2024},
  month = mar,
  journal = {The Review of Economics and Statistics},
  volume = {106},
  number = {2},
  pages = {557--565},
  issn = {0034-6535},
  doi = {10.1162/rest_a_01160},
  urldate = {2024-10-30},
}

@article{mccoy2018polarization,
  title = {Polarization and the Global Crisis of Democracy: Common Patterns, Dynamics, and Pernicious Consequences for Democratic Polities},
  shorttitle = {Polarization and the Global Crisis of Democracy},
  author = {McCoy, Jennifer and Rahman, Tahmina and Somer, Murat},
  year = {2018},
  month = jan,
  journal = {American Behavioral Scientist},
  volume = {62},
  number = {1},
  pages = {16--42},
  publisher = {SAGE Publications Inc},
  issn = {0002-7642},
  doi = {10.1177/0002764218759576},
  urldate = {2024-10-30},
  langid = {english}
}

@article{orhan2022relationship,
  title = {The Relationship between Affective Polarization and Democratic Backsliding: Comparative Evidence},
  shorttitle = {The Relationship between Affective Polarization and Democratic Backsliding},
  author = {Orhan, Yunus Emre},
  year = {2022},
  month = may,
  journal = {Democratization},
  volume = {29},
  number = {4},
  pages = {714--735},
  publisher = {Routledge},
  issn = {1351-0347},
  doi = {10.1080/13510347.2021.2008912},
  urldate = {2024-10-30},
}

@article{reiljan2020fear,
  title = {`Fear and Loathing across Party Lines' (Also) in Europe: Affective Polarisation in European Party Systems},
  shorttitle = {`Fear and Loathing across Party Lines' (Also) in Europe},
  author = {Reiljan, Andres},
  year = {2020},
  journal = {European Journal of Political Research},
  volume = {59},
  number = {2},
  pages = {376--396},
  issn = {1475-6765},
  doi = {10.1111/1475-6765.12351},
  urldate = {2024-10-30},
  copyright = {{\copyright} 2019 European Consortium for Political Research},
  langid = {english},
}

@article{baldassarri2008partisans,
  title = {Partisans without Constraint: Political Polarization and Trends in American Public Opinion},
  author = {Baldassarri, Delia and Gelman, Andrew},
  journal = {American Journal of Sociology},
  year = {2008},
  volume = {114},
  number = {2},
  pages = {404-446},
  langid = {english},
}

@article{garimella2017longterm,
  title = {A Long-Term Analysis of Polarization on Twitter},
  author = {Garimella, Venkata Rama Kiran and Weber, Ingmar},
  year = {2017},
  month = may,
  journal = {Proceedings of the International AAAI Conference on Web and Social Media},
  volume = {11},
  number = {1},
  pages = {528--531},
  issn = {2334-0770},
  doi = {10.1609/icwsm.v11i1.14918},
  urldate = {2024-01-15},
  copyright = {Copyright (c) 2021 Proceedings of the International AAAI Conference on Web and Social Media},
  langid = {english},
}

@article{stoll2010elite,
author = {Heather Stoll},
title = {Elite-Level Conflict Salience and Dimensionality in Western Europe: Concepts and Empirical Findings},
journal = {West European Politics},
volume = {33},
number = {3},
pages = {445--473},
year = {2010},
publisher = {Routledge},
doi = {10.1080/01402381003654494},
}

@article{lelkes2016mass,
  title = {Mass Polarization: Manifestations and Measurements},
  shorttitle = {Mass Polarization},
  author = {Lelkes, Yphtach},
  year = {2016},
  month = jan,
  journal = {Public Opin Q},
  volume = {80},
  number = {S1},
  pages = {392--410},
  publisher = {Oxford Academic},
  issn = {0033-362X},
  doi = {10.1093/poq/nfw005},
  urldate = {2020-10-29},
  langid = {english},
}

@article{kozlowski2021issue,
  title = {Issue Alignment and Partisanship in the American Public: Revisiting the `Partisans without Constraint' Thesis},
  shorttitle = {Issue Alignment and Partisanship in the American Public},
  author = {Kozlowski, Austin C. and Murphy, James P.},
  year = {2021},
  month = feb,
  journal = {Social Science Research},
  volume = {94},
  pages = {102498},
  issn = {0049089X},
  doi = {10.1016/j.ssresearch.2020.102498},
  urldate = {2024-10-30},
  langid = {english},
}

@article{finkel2020political,
  title = {Political Sectarianism in America},
  author = {Finkel, Eli J. and Bail, Christopher A. and Cikara, Mina and Ditto, Peter H. and Iyengar, Shanto and Klar, Samara and Mason, Lilliana and McGrath, Mary C. and Nyhan, Brendan and Rand, David G. and Skitka, Linda J. and Tucker, Joshua A. and Van Bavel, Jay J. and Wang, Cynthia S. and Druckman, James N.},
  year = {2020},
  month = oct,
  journal = {Science},
  volume = {370},
  number = {6516},
  pages = {533--536},
  issn = {0036-8075, 1095-9203},
  doi = {10.1126/science.abe1715},
  urldate = {2024-06-05},
  copyright = {http://www.sciencemag.org/about/science-licenses-journal-article-reuse},
  langid = {english},
}

@article{bornschier2010new,
        author = {Simon Bornschier},
        title = {The New Cultural Divide and the Two-Dimensional Political Space in Western Europe},
        journal = {West European Politics},
        volume = {33},
        number = {3},
        pages = {419--444},
        year = {2010},
        publisher = {Routledge},
        doi = {10.1080/01402381003654387},
}

@article{dolezal2013structure,
        author = {Martin Dolezal, Nikolaus Eder, Sylvia Kritzinger and Eva Zeglovits},
        title = {The Structure of Issue Attitudes Revisited: A Dimensional Analysis of {Austrian} Voters and Party Elites},
        journal = {Journal of Elections, Public Opinion and Parties},
        volume = {23},
        number = {4},
        pages = {423--443},
        year = {2013},
        publisher = {Routledge},
        doi = {10.1080/17457289.2013.803195},
}

@article{fiorina2008political,
  title={Political polarization in the American public},
  author={Fiorina, Morris P and Abrams, Samuel J},
  journal={Annu. Rev. Polit. Sci.},
  volume={11},
  pages={563--588},
  year={2008},
  publisher={Annual Reviews}
}

@article{iyengar2019origins,
  title = {The Origins and Consequences of Affective Polarization in the United States},
  author = {Iyengar, Shanto and Lelkes, Yphtach and Levendusky, Matthew and Malhotra, Neil and Westwood, Sean J.},
  year = {2019},
  month = may,
  journal = {Annual Review of Political Science},
  volume = {22},
  number = {1},
  pages = {129--146},
  issn = {1094-2939, 1545-1577},
  doi = {10.1146/annurev-polisci-051117-073034},
  urldate = {2024-06-27},
  langid = {english},
}

@article{druckman2021affective,
  title={Affective polarization, local contexts and public opinion in America},
  author={Druckman, James N and Klar, Samara and Krupnikov, Yanna and Levendusky, Matthew and Ryan, John Barry},
  journal={Nature human behaviour},
  volume={5},
  number={1},
  pages={28--38},
  year={2021},
  publisher={Nature Publishing Group}
}

@inproceedings{conover2011political,
  title={Political polarization on twitter},
  author={Conover, Michael and Ratkiewicz, Jacob and Francisco, Matthew and Gon{\c{c}}alves, Bruno and Menczer, Filippo and Flammini, Alessandro},
  booktitle={Proceedings of the international aaai conference on web and social media},
  volume={5},
  pages={89--96},
  year={2011},
doi={10.1609/icwsm.v5i1.14126}
}

@article{pham2021balance,
  title = {Balance and Fragmentation in Societies with Homophily and Social Balance},
  author = {Pham, Tuan M. and Alexander, Andrew C. and Korbel, Jan and Hanel, Rudolf and Thurner, Stefan},
  year = {2021},
  month = dec,
  journal = {Scientific Reports},
  volume = {11},
  number = {1},
  pages = {17188},
  issn = {2045-2322},
  doi = {10.1038/s41598-021-96065-5},
  urldate = {2022-11-25},
  langid = {english},
}

@article{baumann2021emergence,
  title = {Emergence of Polarized Ideological Opinions in Multidimensional Topic Spaces},
  author = {Baumann, Fabian and Lorenz-Spreen, Philipp and Sokolov, Igor M. and Starnini, Michele},
  journal = {Phys. Rev. X},
  volume = {11},
  issue = {1},
  pages = {011012},
  numpages = {16},
  year = {2021},
  month = {Jan},
  publisher = {American Physical Society},
  doi = {10.1103/PhysRevX.11.011012},
  url = {https://link.aps.org/doi/10.1103/PhysRevX.11.011012}
}

@article{ches2023,
  author    = {Liesbet Hooghe and Gary Marks and Ryan Bakker and Seth Jolly and Jon Polk and Jan Rovny and Marco Steenbergen and Milada Anna Vachudova},
  title     = {The Russian Threat and the Consolidation of the West: How Populism and {EU}-skepticism shape party support for {Ukraine}},
  journal   = {European Union Politics},
  year      = {2024},
  volume    = {25},
  number    = {3},
  url       = {https://www.chesdata.eu/ches-europe}
}

@article{ches2019,
    title = {Chapel Hill Expert Survey trend file, 1999-2019},
    journal = {Electoral Studies},
    volume = {75},
    pages = {102420},
    year = {2022},
    issn = {0261-3794},
    doi = {10.1016/j.electstud.2021.102420},
    author = {Seth Jolly and Ryan Bakker and Liesbet Hooghe and Gary Marks and Jonathan Polk and Jan Rovny and Marco Steenbergen and Milada Anna Vachudova},
}

@article{hoerl1970ridge,
        author = {Arthur E. Hoerl and Robert W. Kennard},
        title = {Ridge Regression: Biased Estimation for Nonorthogonal Problems},
        journal = {Technometrics},
        volume = {12},
        number = {1},
        pages = {55--67},
        year = {1970},
        publisher = {ASA Website},
        doi = {10.1080/00401706.1970.10488634},
}

@article{bakker2012exploring,
    author = {Ryan Bakker and Seth Jolly and Jonathan Polk},
    title ={Complexity in the {European} party space: Exploring dimensionality with experts},
    journal = {European Union Politics},
    volume = {13},
    number = {2},
    pages = {219-245},
    year = {2012},
    doi = {10.1177/1465116512436995},
}

@article{bakshy2015exposure,
    author = {Eytan Bakshy  and Solomon Messing  and Lada A. Adamic },
    title = {Exposure to ideologically diverse news and opinion on Facebook},
    journal = {Science},
    volume = {348},
    number = {6239},
    pages = {1130-1132},
    year = {2015},
    doi = {10.1126/science.aaa1160},
    URL = {https://www.science.org/doi/abs/10.1126/science.aaa1160},
    eprint = {https://www.science.org/doi/pdf/10.1126/science.aaa1160},
}

@article{bakshy2015exposure_data,
    author = {Eytan Bakshy  and Solomon Messing  and Lada A. Adamic },
    title = {Exposure to ideologically diverse news and opinion on Facebook},
  year = 2015,
  howpublished = {\emph{Harvard Dataverse} \url{https://doi.org/10.7910/DVN/AAI7VA}}
}

@inproceedings{fraisier2018elysee,
  title     = {{\#Élysée2017fr: The 2017 French Presidential Campaign on Twitter}},
  booktitle = {Proceedings of the 12th International AAAI Conference on Web and Social Media},
  author    = {Fraisier, Ophélie and Cabanac, Guillaume and Pitarch, Yoann and Besançon, Romaric and Boughanem, Mohand},
  year      = {2018},
    doi = {10.1609/icwsm.v12i1.14984},
}

@dataset{fraisier2018elysee_data,
  author    = {Fraisier, Ophélie and Cabanac, Guillaume and Pitarch, Yoann and Besançon, Romaric and Boughanem, Mohand},
  title     = {\#Élysée2017fr: The 2017 {French} Presidential Campaign on {Twitter} [dataset]},
  month     = jun,
  year      = 2018,
  howpublished = {\emph{Zenodo} \url{https://doi.org/10.5281/zenodo.5535333}}
}

@article{torcal2023dynamics,
  title = {The Dynamics of Political and Affective Polarisation: {Datasets for Spain, Portugal, Italy, Argentina, and Chile} (2019-2022)},
  shorttitle = {The Dynamics of Political and Affective Polarisation},
  author = {Torcal, Mariano and Carty, Emily and Comellas, Josep Maria and Bosch, Oriol J. and Thomson, Zoe and Serani, Danilo},
  year = {2023},
  month = jun,
  journal = {Data in Brief},
  volume = {48},
  pages = {109219},
  issn = {23523409},
  doi = {10.1016/j.dib.2023.109219},
  urldate = {2024-10-24},
  langid = {english},
}

@dataset{torcal2023dynamics_data,
  author = {Torcal, Mariano and Carty, Emily and Comellas, Josep Maria and Bosch, Oriol J. and Thomson, Zoe and Serani, Danilo},
  title = {The Dynamics of Political and Affective Polarisation: {Datasets for Spain, Portugal, Italy, Argentina, and Chile} (2019-2022) [dataset]},
    year = {2023},
  howpublished = {\emph{OSF} \url{http://doi.org/10.17605/OSF.IO/3T7JZ}}
}

@inproceedings{adamic2005political,
    author = {Adamic, Lada A. and Glance, Natalie},
    title = {The Political Blogosphere and the 2004 {U.S.} Election: {Divided} They Blog},
    year = {2005},
    isbn = {1595932151},
    publisher = {Association for Computing Machinery},
    address = {New York, NY, USA},
    doi = {10.1145/1134271.1134277},
    booktitle = {Proceedings of the 3rd International Workshop on Link Discovery},
    pages = {36-43},
    numpages = {8},
    location = {Chicago, Illinois},
    series = {LinkKDD '05}
}

@dataset{adamic2005political_data,
    author = {Adamic, Lada A. and Glance, Natalie},
    title = {The Political Blogosphere and the 2004 {U.S.} Election: {Divided} They Blog [dataset]},
    year = {2005},
    note = {Non-official repository},
  howpublished = {\emph{Figshare} \url{https://figshare.com/articles/dataset/A_directed_network_of_hyperlinks_between_weblogs_on_US_politics_recorded_in_2005_by_Adamic_and_Glance_/1149954?file=1648333}}
}

@misc{srivastava2021poliwam,
      title={{PoliWAM}: {An} Exploration of a Large Scale Corpus of Political Discussions on {WhatsApp} {Messenger}}, 
      author={Vivek Srivastava and Mayank Singh},
      year={2021},
      eprint={https://arxiv.org/abs/2010.13263}, 
      archivePrefix={arXiv},
      primaryClass={cs.SI},
      url={https://arxiv.org/abs/2010.13263}, 
}

@article{kubin2021role,
  title = {The Role of (Social) Media in Political Polarization: A Systematic Review},
  shorttitle = {The Role of (Social) Media in Political Polarization},
  author = {Kubin, Emily and Von Sikorski, Christian},
  year = {2021},
  month = jul,
  journal = {Annals of the International Communication Association},
  volume = {45},
  number = {3},
  pages = {188--206},
  issn = {2380-8985, 2380-8977},
  doi = {10.1080/23808985.2021.1976070},
  urldate = {2024-10-15},
  langid = {english},
}

@dataset{srivastava2021poliwam_data,
      title={{PoliWAM}: {An} Exploration of a Large Scale Corpus of Political Discussions on {WhatsApp} {Messenger} [dataset]}, 
      author={Vivek Srivastava and Mayank Singh},
      year={2021},
      howpublished = {\emph{Zenodo} \url{https://zenodo.org/records/4115660#.X5BHSngzZQI}}
}

@misc{fan2020englishcentric,
      title={Beyond English-Centric Multilingual Machine Translation}, 
      author={Angela Fan and Shruti Bhosale and Holger Schwenk and Zhiyi Ma and Ahmed El-Kishky and Siddharth Goyal and Mandeep Baines and Onur Celebi and Guillaume Wenzek and Vishrav Chaudhary and Naman Goyal and Tom Birch and Vitaliy Liptchinsky and Sergey Edunov and Edouard Grave and Michael Auli and Armand Joulin},
      year={2020},
      eprint={2010.11125},
      archivePrefix={arXiv},
      primaryClass={cs.CL}
}

@misc{tunstall2023zephyr,
      title={Zephyr: Direct Distillation of LM Alignment}, 
      author={Lewis Tunstall and Edward Beeching and Nathan Lambert and Nazneen Rajani and Kashif Rasul and Younes Belkada and Shengyi Huang and Leandro von Werra and Clémentine Fourrier and Nathan Habib and Nathan Sarrazin and Omar Sanseviero and Alexander M. Rush and Thomas Wolf},
      year={2023},
      eprint={2310.16944},
      archivePrefix={arXiv},
      primaryClass={cs.LG}
}

@article{heseltine2024large,
author = {Michael Heseltine and Bernhard Clemm von Hohenberg},
title ={Large language models as a substitute for human experts in annotating political text},
journal = {Research \& Politics},
volume = {11},
number = {1},
pages = {20531680241236239},
year = {2024},
doi = {10.1177/20531680241236239}
}

@article{ramaciotti2024attitudinally,
  title={Attitudinally-positioned European sample dataset},
  author={Ramaciotti, Pedro and Royo-Letelier, Jimena and Cointet, Jean-Philippe and Pournaki, Armin},
  year={2024},
  journal={Report.}
}

@article{mens2024positioning, 
    title={Positioning Political Texts with Large Language Models by Asking and Averaging}, 
    DOI={10.1017/pan.2024.29}, 
    journal={Political Analysis}, 
    author={Le Mens, Gaël and Gallego, Aina}, 
    year={2025}, 
    pages={1–9}
}

@article{clopper1934theuse,
 ISSN = {00063444, 14643510},
 doi = {10.2307/2331986},
 author = {C. J. Clopper and E. S. Pearson},
 journal = {Biometrika},
 number = {4},
 pages = {404--413},
 publisher = {Oxford University Press, Biometrika Trust},
 title = {The Use of Confidence or Fiducial Limits Illustrated in the Case of the Binomial},
 urldate = {2025-02-03},
 volume = {26},
 year = {1934}
}

@unpublished{corpora_paper,
      title={Uncovering the structure of the {French} media ecosystem}, 
      author={Jean-Philippe Cointet and Dominique Cardon and Andreï Mogoutov and Benjamin Ooghe-Tabanou and Guillaume Plique and Pedro Morales},
      year={2021},
      eprint={2107.12073},
      archivePrefix={arXiv},
      primaryClass={cs.SI},
      url={https://arxiv.org/abs/2107.12073}, 
}

@dataset{corpora_data,
author = {Plique, Guillaume and Ooghe-Tabanou, Benjamin and Cointet, Jean-Philippe and Cardon, Dominique and Baneyx, Audrey and Girard, Paul and Pichon, Arnaud and Crepel, Maxime and Perrin, Oubine and Antolinos-Basso, Diego and Tainturier, Benjamin and Fingerhut, Tim},
publisher = {data.sciencespo},
title = {{Corpus Médias ``Polarisation de l'Espace Public Numérique''} [dataset]},
UNF = {UNF:6:w9rjfXKW6yUoBRZ8XXTC+A==},
year = {2021},
version = {V3},
howpublished = {\emph{data.sciencespo} \url{https://doi.org/10.21410/7E4/HZB8D0}}
}

@article{rottger2021two,
  title={Two contrasting data annotation paradigms for subjective NLP tasks},
  author={R{\"o}ttger, Paul and Vidgen, Bertie and Hovy, Dirk and Pierrehumbert, Janet B},
  journal={arXiv preprint arXiv:2112.07475},
  year={2021}
}

@dataset{defacto_data,
    title = {French Media Ecosystem Map [dataset]},
    author = {Crépel, Maxime and Ooghe-Tabanou, Benjamin and Plique, Guillaume and Mazoyer, Béatrice and Christensen, Kelly and Cointet, Jean-Philippe and Parasie, Sylvain and Cardon, Dominique and Machut, Antoine and Tittel, Katharina},
    urldate = {2024-11-05},
    year =  {2024},
    howpublished ={\emph{SciencesPo data} \url{https://doi.org/10.21410/7E4/VMMY7L}}
}

@inproceedings{ramaciotti2022measuring,
  title={Measuring the accuracy of social network ideological embeddings using language models},
  author={Ramaciotti Morales, Pedro and Mu{\~n}oz Zolotoochin, Gabriel},
  booktitle={International Conference on Information Technology \& Systems},
  pages={267--276},
  year={2022},
  organization={Springer}
}

@techreport{defacto_paper,
  TITLE = {{Digital mapping of the French media ecosystem}},
  AUTHOR = {Cr{\'e}pel, Maxime and Ooghe-Tabanou, Benjamin and Christensen, Kelly and Mazoyer, B{\'e}atrice and Plique, Guillaume and Parasie, Sylvain and Cardon, Dominique and Tittel, Katharina and Machut, Antoine and Cointet, Jean-Pilippe},
  URL = {https://sciencespo.hal.science/hal-04868603},
  NUMBER = {Sciences Po m{\'e}dialab - DEFACTO - 02/23/24},
  PUBLISHER = {{data.sciencespo}},
  PAGES = {13 p.},
  INSTITUTION = {{M{\'e}dialab ; AFP ; CLEMI ; Wiki}},
  YEAR = {2024},
  HAL_ID = {hal-04868603},
  HAL_VERSION = {v1},
}

@article{morstatter2021sample,
  title = {Is the Sample Good Enough? {Comparing} Data from {Twitter}'s Streaming {API} with {Twitter}'s Firehose},
  shorttitle = {Is the Sample Good Enough?},
  author = {Morstatter, Fred and Pfeffer, J{\"u}rgen and Liu, Huan and Carley, Kathleen},
  year = {2021},
  month = aug,
  journal = {Proceedings of the International AAAI Conference on Web and Social Media},
  volume = {7},
  number = {1},
  pages = {400--408},
  issn = {2334-0770, 2162-3449},
  doi = {10.1609/icwsm.v7i1.14401},
  urldate = {2024-06-25},
  langid = {english},
}

@misc{dibona2024sampled,
      title={Sampled Datasets Risk Substantial Bias in the Identification of Political Polarization on Social Media}, 
      author={Gabriele Di Bona and Emma Fraxanet and Björn Komander and Andrea Lo Sasso and Virginia Morini and Antoine Vendeville and Max Falkenberg and Alessandro Galeazzi},
      year={2024},
      eprint={https://arxiv.org/abs/2406.19867},
      archivePrefix={arXiv},
      primaryClass={cs.SI},
      url={https://arxiv.org/abs/2406.19867}, 
}

@article{falkenberg2024patterns,
  title = {Patterns of Partisan Toxicity and Engagement Reveal the Common Structure of Online Political Communication across Countries},
  author = {Falkenberg, Max and Zollo, Fabiana and Quattrociocchi, Walter and Pfeffer, J{\"u}rgen and Baronchelli, Andrea},
  year = {2024},
  month = nov,
  journal = {Nat Commun},
  volume = {15},
  number = {1},
  pages = {9560},
  issn = {2041-1723},
  doi = {10.1038/s41467-024-53868-0},
  urldate = {2024-11-18},
  langid = {english},
}

@article{clinton2004statistical,
  title={The statistical analysis of roll call data},
  author={Clinton, Joshua and Jackman, Simon and Rivers, Douglas},
  journal={American Political Science Review},
  volume={98},
  number={2},
  pages={355--370},
  year={2004},
  publisher={Cambridge University Press}
}

@article{luskin1990explaining,
  title={Explaining political sophistication},
  author={Luskin, Robert C},
  journal={Political behavior},
  volume={12},
  pages={331--361},
  year={1990},
  publisher={Springer}
}

@article{flamino2023political,
  title={Political polarization of news media and influencers on {Twitter} in the 2016 and 2020 {US} presidential elections},
  author={Flamino, James and Galeazzi, Alessandro and Feldman, Stuart and Macy, Michael W and Cross, Brendan and Zhou, Zhenkun and Serafino, Matteo and Bovet, Alexandre and Makse, Hern{\'a}n A and Szymanski, Boleslaw K},
  journal={Nature Human Behaviour},
  volume={7},
  number={6},
  pages={904--916},
  year={2023},
  publisher={Nature Publishing Group UK London}
}

@misc{ess2023,
  author       = {{European Social Survey European Research Infrastructure (ESS ERIC)}},
  title        = {{ESS Round 11 - 2023: Social Inequalities in Health, Gender in Contemporary Europe}},
  year         = {2023},
  publisher    = {Sikt - Norwegian Agency for Shared Services in Education and Research},
  doi          = {10.21338/ess11-2023},
  url          = {https://doi.org/10.21338/ess11-2023}
}

@article{Mellon2017,
  title = {Twitter and Facebook Are Not Representative of the General Population: Political Attitudes and Demographics of British Social Media Users},
  shorttitle = {Twitter and Facebook Are Not Representative of the General Population},
  author = {Mellon, Jonathan and Prosser, Christopher},
  year = {2017},
  month = jul,
  journal = {Research \& Politics},
  volume = {4},
  number = {3},
  pages = {2053168017720008},
  issn = {2053-1680, 2053-1680},
  doi = {10.1177/2053168017720008},
  urldate = {2025-07-03},
  langid = {english},
}

@article{Zagheni2015,
  title = {Demographic Research with Non-Representative Internet Data},
  author = {Zagheni, Emilio and Weber, Ingmar},
  editor = {Nikolaos Askitas And Professor Klaus F. Zimmermann, Professor},
  year = {2015},
  month = apr,
  journal = {International Journal of Manpower},
  volume = {36},
  number = {1},
  pages = {13--25},
  issn = {0143-7720},
  doi = {10.1108/IJM-12-2014-0261},
  urldate = {2025-07-03},
  copyright = {https://www.emerald.com/insight/site-policies},
  langid = {english},
}

@article{falkenberg2022growing,
  title = {Growing Polarization around Climate Change on Social Media},
  author = {Falkenberg, Max and Galeazzi, Alessandro and Torricelli, Maddalena and Di Marco, Niccol{\`o} and Larosa, Francesca and Sas, Madalina and Mekacher, Amin and Pearce, Warren and Zollo, Fabiana and Quattrociocchi, Walter and Baronchelli, Andrea},
  year = {2022},
  month = dec,
  journal = {Nature Climate Change},
  volume = {12},
  number = {12},
  pages = {1114--1121},
  issn = {1758-678X, 1758-6798},
  doi = {10.1038/s41558-022-01527-x},
  urldate = {2024-03-28},
  langid = {english},
}

@article{tromble2021where,
    author = {Rebekah Tromble},
    title ={Where Have All the Data Gone? A Critical Reflection on Academic Digital Research in the Post-API Age},
    journal = {Social Media + Society},
    volume = {7},
    number = {1},
    pages = {2056305121988929},
    year = {2021},
    doi = {10.1177/2056305121988929},
}

@article{ozkula2023easy,
  title = {Easy Data, Same Old Platforms? {A} Systematic Review of Digital Activism Methodologies},
  shorttitle = {Easy Data, Same Old Platforms?},
  author = {{\"O}zkula, Suay M. and Reilly, Paul J. and Hayes, Jenny},
  year = {2023},
  month = may,
  journal = {Information, Communication \& Society},
  volume = {26},
  number = {7},
  pages = {1470--1489},
  issn = {1369-118X, 1468-4462},
  doi = {10.1080/1369118X.2021.2013918},
  urldate = {2024-10-10},
  langid = {english},
}

@article{roozenbeek2022democratize,
  title={Democratize social-media research-with access and funding},
  author={Roozenbeek, Jon and Zollo, Fabiana},
  journal={Nature},
  volume={612},
  number={7940},
  pages={404},
  year={2022},
doi = {10.1038/d41586-022-04407-8}
}

@book{greenacre2017correspondence,
  author       = {Greenacre, Michael},
  title        = {Correspondence Analysis in Practice},
  year         = {2017},
  edition      = {Third Edition},
  publisher    = {CRC Press},
  address      = {Boca Raton, FL},
  isbn         = {9781315369983},
  doi          = {10.1201/9781315369983},
}

@article{caroll1997equivalence,
author = {Carroll, J. Douglas and Kumbasar, Ece and Romney, A. Kimball},
title = {An equivalence relation between correspondence analysis and classical metric multidimensional scaling for the recovery of Euclidean distances},
journal = {British Journal of Mathematical and Statistical Psychology},
volume = {50},
number = {1},
pages = {81-92},
doi = {10.1111/j.2044-8317.1997.tb01104.x},
year = {1997}
}

@article{lowe2008understanding, 
    title={Understanding Wordscores}, 
    volume={16}, 
    DOI={10.1093/pan/mpn004}, 
    number={4}, 
    journal={Political Analysis}, 
    author={Lowe, Will}, 
    year={2008}, 
    pages={356–371}
}

@article{bond2015quantifying, 
    title={Quantifying Social Media’s Political Space: Estimating Ideology from Publicly Revealed Preferences on Facebook}, 
    volume={109}, 
    DOI={10.1017/S0003055414000525}, 
    number={1}, 
    journal={American Political Science Review}, 
    author={Bond, Robert and Messing, Solomon}, 
    year={2015}, 
    pages={62–78}
}

@INPROCEEDINGS{ramaciotti2020your,
  author={Morales, Pedro Ramaciotti and Cointet, Jean-Philippe and Laborde, Julio},
  booktitle={2020 IEEE/ACM International Conference on Advances in Social Networks Analysis and Mining (ASONAM)}, 
  title={Your most telling friends: Propagating latent ideological features on Twitter using neighborhood coherence}, 
  year={2020},
  volume={},
  number={},
  pages={217-221},
  doi={10.1109/ASONAM49781.2020.9381468}
}

@inproceedings{briatte2015recovering,
  TITLE = {Recovering the {French} Party Space from {Twitter} Data},
  AUTHOR = {Briatte, François and Gallic, Ewen},
  URL = {https://shs.hal.science/halshs-01511384},
  BOOKTITLE = {{Science Po Quanti}},
  ADDRESS = {Paris, France},
  YEAR = {2015},
  MONTH = May,
  HAL_ID = {halshs-01511384},
  HAL_VERSION = {v1},
}

@article{barbera2015understanding,
    author = {Pablo Barberá and Gonzalo Rivero},
    title ={Understanding the Political Representativeness of {Twitter} Users},
    journal = {Social Science Computer Review},
    volume = {33},
    number = {6},
    pages = {712-729},
    year = {2015},
}

@article{barbera2015birds,
  title = {Birds of the Same Feather Tweet Together: Bayesian Ideal Point Estimation Using {Twitter} Data},
  shorttitle = {Birds of the Same Feather Tweet Together},
  author = {Barber{\'a}, Pablo},
  year = {2015},
  journal = {Political Analysis},
  volume = {23},
  number = {1},
  pages = {76--91},
  publisher = {Cambridge University Press},
  issn = {1047-1987, 1476-4989},
  doi = {10.1093/pan/mpu011},
  urldate = {2020-11-03},
  langid = {english},
}

@inproceedings{corso2024what,
author = {Corso, Francesco and Pierri, Francesco and De Francisci Morales, Gianmarco},
title = {What we can learn from {TikTok} through its Research {API}},
year = {2024},
isbn = {9798400704536},
publisher = {Association for Computing Machinery},
address = {New York, NY, USA},
doi = {10.1145/3630744.3663611},
booktitle = {Companion Publication of the 16th ACM Web Science Conference},
pages = {110–114},
numpages = {5},
location = {Stuttgart, Germany},
series = {Websci Companion '24}
}

@inproceedings{papasavva2021voat,
author = {Papasavva, Antonis and Blackburn, Jeremy and Stringhini, Gianluca and Zannettou, Savvas and Cristofaro, Emiliano De},
title = {“{Is it a Qoincidence}?”: An Exploratory Study of {QAnon} on {Voat}},
year = {2021},
isbn = {9781450383127},
publisher = {Association for Computing Machinery},
address = {New York, NY, USA},
doi = {10.1145/3442381.3450036},
booktitle = {Proceedings of the Web Conference 2021},
pages = {460–471},
numpages = {12},
location = {Ljubljana, Slovenia},
series = {WWW '21}
}

@article{papasavva2020raiders, 
    title={Raiders of the {Lost} {Kek}: 3.5 Years of Augmented 4chan Posts from the Politically Incorrect Board}, 
    volume={14}, url={https://ojs.aaai.org/index.php/ICWSM/article/view/7354}, 
    DOI={https://doi.org/10.1609/icwsm.v14i1.7354}, 
    number={1}, 
    journal={Proceedings of the International AAAI Conference on Web and Social Media}, 
    author={Papasavva, Antonis and Zannettou, Savvas and De Cristofaro, Emiliano and Stringhini, Gianluca and Blackburn, Jeremy}, 
    year={2020}, 
    month={May}, 
    pages={885-894} 
}

@article{gonzalezbailon2023asymmetric,
  title = {Asymmetric Ideological Segregation in Exposure to Political News on {Facebook}},
  author = {{Gonz{\'a}lez-Bail{\'o}n}, Sandra and Lazer, David and Barber{\'a}, Pablo and Zhang, Meiqing and Allcott, Hunt and Brown, Taylor and {Crespo-Tenorio}, Adriana and Freelon, Deen and Gentzkow, Matthew and Guess, Andrew M. and Iyengar, Shanto and Kim, Young Mie and Malhotra, Neil and Moehler, Devra and Nyhan, Brendan and Pan, Jennifer and Rivera, Carlos Velasco and Settle, Jaime and Thorson, Emily and Tromble, Rebekah and Wilkins, Arjun and Wojcieszak, Magdalena and De Jonge, Chad Kiewiet and Franco, Annie and Mason, Winter and Stroud, Natalie Jomini and Tucker, Joshua A.},
  year = {2023},
  month = jul,
  journal = {Science},
  volume = {381},
  number = {6656},
  pages = {392--398},
  issn = {0036-8075, 1095-9203},
  doi = {10.1126/science.ade7138},
  urldate = {2023-12-02},
  langid = {english},
}

@article{hartigan1985dip,
  title={The dip test of unimodality},
  author={Hartigan, John A and Hartigan, Pamela M},
  journal={The annals of Statistics},
  pages={70--84},
  year={1985},
  publisher={JSTOR}
}

@article{guess2023reshares,
  title = {Reshares on Social Media Amplify Political News but Do Not Detectably Affect Beliefs or Opinions},
  author = {Guess, Andrew M. and Malhotra, Neil and Pan, Jennifer and Barber{\'a}, Pablo and Allcott, Hunt and Brown, Taylor and {Crespo-Tenorio}, Adriana and Dimmery, Drew and Freelon, Deen and Gentzkow, Matthew and {Gonz{\'a}lez-Bail{\'o}n}, Sandra and Kennedy, Edward and Kim, Young Mie and Lazer, David and Moehler, Devra and Nyhan, Brendan and Rivera, Carlos Velasco and Settle, Jaime and Thomas, Daniel Robert and Thorson, Emily and Tromble, Rebekah and Wilkins, Arjun and Wojcieszak, Magdalena and Xiong, Beixian and {de Jonge}, Chad Kiewiet and Franco, Annie and Mason, Winter and Stroud, Natalie Jomini and Tucker, Joshua A.},
  year = {2023},
  month = jul,
  journal = {Science},
  volume = {381},
  number = {6656},
  pages = {404--408},
  publisher = {American Association for the Advancement of Science},
  doi = {10.1126/science.add8424},
  urldate = {2023-07-28},
}

@article{guess2023howdo,
  title = {How Do Social Media Feed Algorithms Affect Attitudes and Behavior in an Election Campaign?},
  author = {Guess, Andrew M. and Malhotra, Neil and Pan, Jennifer and Barber{\'a}, Pablo and Allcott, Hunt and Brown, Taylor and {Crespo-Tenorio}, Adriana and Dimmery, Drew and Freelon, Deen and Gentzkow, Matthew and {Gonz{\'a}lez-Bail{\'o}n}, Sandra and Kennedy, Edward and Kim, Young Mie and Lazer, David and Moehler, Devra and Nyhan, Brendan and Rivera, Carlos Velasco and Settle, Jaime and Thomas, Daniel Robert and Thorson, Emily and Tromble, Rebekah and Wilkins, Arjun and Wojcieszak, Magdalena and Xiong, Beixian and {de Jonge}, Chad Kiewiet and Franco, Annie and Mason, Winter and Stroud, Natalie Jomini and Tucker, Joshua A.},
  year = {2023},
  month = jul,
  journal = {Science},
  volume = {381},
  number = {6656},
  pages = {398--404},
  publisher = {American Association for the Advancement of Science},
  doi = {10.1126/science.abp9364},
  urldate = {2023-07-28},
}

@article{nyhan2023likeminded,
  title = {Like-Minded Sources on {Facebook} Are Prevalent but Not Polarizing},
  author = {Nyhan, Brendan and Settle, Jaime and Thorson, Emily and Wojcieszak, Magdalena and Barber{\'a}, Pablo and Chen, Annie Y. and Allcott, Hunt and Brown, Taylor and {Crespo-Tenorio}, Adriana and Dimmery, Drew and Freelon, Deen and Gentzkow, Matthew and {Gonz{\'a}lez-Bail{\'o}n}, Sandra and Guess, Andrew M. and Kennedy, Edward and Kim, Young Mie and Lazer, David and Malhotra, Neil and Moehler, Devra and Pan, Jennifer and Thomas, Daniel Robert and Tromble, Rebekah and Rivera, Carlos Velasco and Wilkins, Arjun and Xiong, Beixian and De Jonge, Chad Kiewiet and Franco, Annie and Mason, Winter and Stroud, Natalie Jomini and Tucker, Joshua A.},
  year = {2023},
  month = aug,
  journal = {Nature},
  volume = {620},
  number = {7972},
  pages = {137--144},
  issn = {0028-0836, 1476-4687},
  doi = {10.1038/s41586-023-06297-w},
  urldate = {2023-12-02},
  langid = {english},
}

@article{mekacher2023systemic,
  title = {The Systemic Impact of Deplatforming on Social Media},
  author = {Mekacher, Amin and Falkenberg, Max and Baronchelli, Andrea},
  year = {2023},
  month = nov,
  journal = {PNAS Nexus},
  volume = {2},
  number = {11},
  pages = {pgad346},
  issn = {2752-6542},
  doi = {10.1093/pnasnexus/pgad346},
  urldate = {2024-06-27},
}

@article{mekacher2024koo,
  title = {The {Koo} Dataset: An {Indian} Microblogging Platform with Global Ambitions},
  shorttitle = {The Koo Dataset},
  author = {Mekacher, Amin and Falkenberg, Max and Baronchelli, Andrea},
  year = {2024},
  month = may,
  journal = {Proceedings of the International AAAI Conference on Web and Social Media},
  volume = {18},
  pages = {1991--2002},
  issn = {2334-0770},
  doi = {10.1609/icwsm.v18i1.31442},
  urldate = {2024-06-27},
  copyright = {Copyright (c) 2024 Association for the Advancement of Artificial Intelligence},
  langid = {english},
}

@article{wagner2023independence,
    author = {Michael W. Wagner},
    title = {Independence by permission},
    journal = {Science},
    volume = {381},
    number = {6656},
    pages = {388-391},
    year = {2023},
    doi = {10.1126/science.adi2430},
}

@article{peralta2024multidimensional,
  title = {Multidimensional political polarization in online social networks},
  author = {Peralta, Antonio F. and Ramaciotti, Pedro and Kert\'esz, J\'anos and I\~niguez, Gerardo},
  journal = {Phys. Rev. Res.},
  volume = {6},
  issue = {1},
  pages = {013170},
  numpages = {10},
  year = {2024},
  month = {Feb},
  publisher = {American Physical Society},
  doi = {10.1103/PhysRevResearch.6.013170},
}

@article{ramaciotti2022inferring,
  title = {Inferring Attitudinal Spaces in Social Networks},
  author = {Ramaciotti Morales, Pedro and Cointet, Jean-Philippe and Mu{\~n}oz Zolotoochin, Gabriel and Fern{\'a}ndez Peralta, Antonio and I{\~n}iguez, Gerardo and Pournaki, Armin},
  year = {2022},
  month = dec,
  journal = {Social Network Analysis and Mining},
  volume = {13},
  number = {1},
  pages = {14},
  issn = {1869-5469},
  doi = {10.1007/s13278-022-01013-4},
  urldate = {2022-12-31},
  langid = {english},
}

@inproceedings{ramaciotti2021unfolding,
author = {Morales, Pedro Ramaciotti and Cointet, Jean-Philippe and Zolotoochin, Gabriel Mu\~{n}oz},
title = {Unfolding the dimensionality structure of social networks in ideological embeddings},
year = {2022},
isbn = {9781450391283},
publisher = {Association for Computing Machinery},
address = {New York, NY, USA},
url = {https://doi.org/10.1145/3487351.3489441},
doi = {10.1145/3487351.3489441},
booktitle = {Proceedings of the 2021 IEEE/ACM International Conference on Advances in Social Networks Analysis and Mining},
pages = {333–338},
numpages = {6},
location = {Virtual Event, Netherlands},
series = {ASONAM '21}
}

@software{ooghe2023gazouilloire,
  author       = {Benjamin Ooghe-Tabanou and
                  Béatrice Mazoyer and
                  Jules Farjas and
                  Guillaume Plique},
  title        = {{Gazouilloire, a command-line tool for long term 
                   collections of tweets}},
  month        = jul,
  year         = 2023,
  publisher    = {Zenodo},
  version      = {v1.5.0},
  doi          = {10.5281/zenodo.8108616},
  url          = {https://doi.org/10.5281/zenodo.8108616}
}

\begin{appendices}
\section*{Appendix}

\section{Impact of retaining users with less than 25 followers}
In a seminal paper \cite{barbera2015birds}, Barbera suggests removing accounts with less than 25 followers before computing political positions, in order to filter out bots and inactive accounts. However, to allow for research projects to investigate the relations between political opinions and activity, we choose to retain these accounts. This changes the structure of the bipartite follow network, and therefore may impact results for all users and MPs. Our full dataset has 980K users and MPs, among which 518K have at least 25 followers. We now demonstrate that our choice does not significantly alter the results. To do so, we create an alternative version of the dataset where political positions are computed solely for users with at least 25 followers. We compare these positions with those obtained for the same users with the full dataset. We find very high Pearson correlations between the two, ranging from 0.909 (dimension: lrecon\_23) to 0.997 (eu\_position\_19), with an average over all dimensions of 0.973. Retaining users with less than 25 followers therefore does not impact the results, justifying our choice to leave them in. 

\section{Parties and political dimensions}
We present the different parties that appear in the data and their characteristics in Table~\ref{tab:parties}. Note the following two differences in nomenclature between our data and the CHES surveys: RE in our data corresponds to LREM in CHES, LFI in our data corresponds to FI in CHES. The following additional parties can be found in our data: Horizons (center right), PRV (Parti Radical, center-right), and LC (Les Centristes, center). MPs without party affiliation are noted as Independent.

\begin{table*}[h!]
\setlength\extrarowheight{2pt} 
\scriptsize
\centering
\caption{Characteristics of the different parties in the dataset. The number of MPs refers to the presence in our data. The last two columns indicate whether or not the party was included in the CHES surveys.}
\begin{tabularx}{\textwidth}{|lXccc|}
\hline
\textbf{Acronym} & \textbf{Name} & \textbf{MPs} & \textbf{CHES 2019} & \textbf{CHES 2023} \\
\hline
RE & Renaissance & 193 & \checkmark & \checkmark \\
LR & Les Républicains & 172 & \checkmark & \checkmark \\
RN & Rassemblement National & 87 & \checkmark & \checkmark \\
PS & Parti Socialiste & 84 & \checkmark & \checkmark \\
LFI & La France Insoumise & 72 & \checkmark & \checkmark \\
MoDem & Mouvement Démocrate & 52 & \checkmark & \checkmark \\
Horizons & Horizons & 36 & - & - \\
UDI & Union des Démocrates et Indépendants & 33 & - & \checkmark \\
EELV & Europe Ecologie Les Verts & 26 & \checkmark & \checkmark \\
PCF & Parti Communiste Français & 25 & \checkmark & \checkmark \\
PRV & Parti Radical &  14 & - & - \\
LC & Les Centristes & 7 & - & - \\
PRG & Parti Radical de Gauche & 2 & - & \checkmark \\
DLF & Debout La France & 1 & \checkmark & \checkmark \\
\hline
\end{tabularx}
\label{tab:parties}
\end{table*}

\section{Number of annotated bios}
Table~\ref{tab:validation_full} provides the number of samples per label, i.e., the number of annotated bios, used for the logistic regression models in the validation, as well as the F1-score of the model. For the sake of space we do not include precision and recall, which are computed in the Jupyter notebook accompanying the data.

\begin{table*}
\setlength\extrarowheight{2pt} 
\tiny
\centering
\caption{Number of annotated bios, F1-score, Precision and Recall of the logistic regression models used for the validation. Sorted by decreasing F1-score.}
\begin{tabularx}{\textwidth}{|lllXXrrrrr|}
\hline
\textbf{Dimension} & \textbf{Year} & \textbf{Annotator} & \textbf{Label A} & \textbf{Label B} & $N_A$ & $N_B$ & \textbf{F1-score} & \textbf{Precision} & \textbf{Recall} \\
\hline
lrgen & 2019 & human & left & right & 1975 & 1593 & 0.965 & 0.976 & 0.953 \\
lrgen & 2019 & human & right & left & 1593 & 1975 & 0.957 & 0.944 & 0.971 \\
lrecon & 2019 & human & left & right & 1975 & 1593 & 0.957 & 0.972 & 0.941 \\
lrecon & 2023 & human & left & right & 1975 & 1593 & 0.951 & 0.977 & 0.926 \\
lrecon & 2019 & human & right & left & 1593 & 1975 & 0.948 & 0.930 & 0.967 \\
lrecon & 2023 & human & right & left & 1593 & 1975 & 0.942 & 0.913 & 0.973 \\
eu\_position & 2023 & human & pro\_european & eurosceptic & 880 & 536 & 0.940 & 0.962 & 0.919 \\
eu\_position & 2019 & human & pro\_european & eurosceptic & 880 & 536 & 0.938 & 0.961 & 0.916 \\
refugees & 2023 & human & restrictive\_immigration & liberal\_immigration & 211 & 161 & 0.933 & 0.946 & 0.919 \\
immigrate\_policy & 2019 & human & restrictive\_immigration & liberal\_immigration & 211 & 161 & 0.933 & 0.946 & 0.919 \\
galtan & 2019 & LLM & liberal & conservative & 9519 & 4666 & 0.920 & 0.939 & 0.901 \\
galtan & 2023 & LLM & liberal & conservative & 9519 & 4666 & 0.919 & 0.940 & 0.898 \\
immigrate\_policy & 2019 & human & liberal\_immigration & restrictive\_immigration & 161 & 211 & 0.915 & 0.898 & 0.932 \\
refugees & 2023 & human & liberal\_immigration & restrictive\_immigration & 161 & 211 & 0.915 & 0.898 & 0.932 \\
sociallifestyle & 2019 & LLM & liberal & conservative & 9519 & 4666 & 0.914 & 0.944 & 0.887 \\
eu\_position & 2023 & human & eurosceptic & pro\_european & 536 & 880 & 0.907 & 0.877 & 0.940 \\
eu\_position & 2019 & human & eurosceptic & pro\_european & 536 & 880 & 0.904 & 0.872 & 0.938 \\
antielite\_salience & 2023 & LLM & elite & populist & 51070 & 5265 & 0.900 & 0.985 & 0.828 \\
eu\_position & 2023 & LLM & pro\_european & eurosceptic & 15846 & 5422 & 0.896 & 0.950 & 0.847 \\
antielite\_salience & 2019 & LLM & elite & populist & 51070 & 5265 & 0.894 & 0.981 & 0.822 \\
immigrate\_policy & 2019 & LLM & restrictive\_immigration & liberal\_immigration & 5262 & 3350 & 0.889 & 0.918 & 0.862 \\
eu\_position & 2019 & LLM & pro\_european & eurosceptic & 15846 & 5422 & 0.880 & 0.949 & 0.820 \\
antielite\_salience & 2023 & human & elite & populist & 416 & 198 & 0.875 & 0.923 & 0.832 \\
nationalism & 2019 & LLM & cosmopolitan & nationalist & 10345 & 9300 & 0.873 & 0.864 & 0.882 \\
antielite\_salience & 2019 & human & elite & populist & 416 & 198 & 0.871 & 0.902 & 0.841 \\
lrgen & 2019 & LLM & left & right & 20477 & 17125 & 0.870 & 0.901 & 0.842 \\
refugees & 2023 & LLM & restrictive\_immigration & liberal\_immigration & 5262 & 3350 & 0.869 & 0.911 & 0.831 \\
lrecon & 2019 & LLM & left & right & 20477 & 17125 & 0.864 & 0.910 & 0.823 \\
lrecon & 2023 & LLM & left & right & 20477 & 17125 & 0.859 & 0.913 & 0.810 \\
lrgen & 2019 & LLM & right & left & 17125 & 20477 & 0.856 & 0.825 & 0.889 \\
nationalism & 2019 & LLM & nationalist & cosmopolitan & 9300 & 10345 & 0.856 & 0.866 & 0.846 \\
lrecon & 2019 & LLM & right & left & 17125 & 20477 & 0.854 & 0.810 & 0.903 \\
lrecon & 2023 & LLM & right & left & 17125 & 20477 & 0.850 & 0.800 & 0.907 \\
galtan & 2019 & LLM & conservative & liberal & 4666 & 9519 & 0.846 & 0.813 & 0.881 \\
galtan & 2023 & LLM & conservative & liberal & 4666 & 9519 & 0.845 & 0.810 & 0.884 \\
corrupt\_salience & 2019 & LLM & elite & populist & 51070 & 5265 & 0.841 & 0.976 & 0.738 \\
sociallifestyle & 2019 & LLM & conservative & liberal & 4666 & 9519 & 0.840 & 0.794 & 0.892 \\
immigrate\_policy & 2019 & LLM & liberal\_immigration & restrictive\_immigration & 3350 & 5262 & 0.839 & 0.802 & 0.879 \\
refugees & 2023 & LLM & liberal\_immigration & restrictive\_immigration & 3350 & 5262 & 0.816 & 0.767 & 0.873 \\
people\_vs\_elite & 2019 & LLM & elite & populist & 51070 & 5265 & 0.812 & 0.961 & 0.704 \\
antielite\_salience & 2023 & human & populist & elite & 198 & 416 & 0.773 & 0.707 & 0.854 \\
antielite\_salience & 2019 & human & populist & elite & 198 & 416 & 0.755 & 0.708 & 0.808 \\
eu\_position & 2023 & LLM & eurosceptic & pro\_european & 5422 & 15846 & 0.751 & 0.660 & 0.870 \\
eu\_position & 2019 & LLM & eurosceptic & pro\_european & 5422 & 15846 & 0.726 & 0.623 & 0.870 \\
people\_vs\_elite & 2019 & human & elite & populist & 416 & 198 & 0.722 & 0.818 & 0.647 \\
environment & 2019 & LLM & pro\_environment & climate\_denialist & 28163 & 107 & 0.714 & 0.997 & 0.556 \\
corrupt\_salience & 2019 & human & elite & populist & 416 & 198 & 0.699 & 0.836 & 0.601 \\
environment & 2019 & LLM & pro\_environment & economic\_focus & 28163 & 8387 & 0.661 & 0.858 & 0.538 \\
corrupt\_salience & 2019 & human & populist & elite & 198 & 416 & 0.581 & 0.473 & 0.753 \\
people\_vs\_elite & 2019 & human & populist & elite & 198 & 416 & 0.571 & 0.484 & 0.697 \\
antielite\_salience & 2023 & LLM & populist & elite & 5265 & 51070 & 0.495 & 0.345 & 0.878 \\
antielite\_salience & 2019 & LLM & populist & elite & 5265 & 51070 & 0.474 & 0.329 & 0.848 \\
environment & 2019 & LLM & economic\_focus & pro\_environment & 8387 & 28163 & 0.431 & 0.311 & 0.700 \\
corrupt\_salience & 2019 & LLM & populist & elite & 5265 & 51070 & 0.378 & 0.245 & 0.825 \\
people\_vs\_elite & 2019 & LLM & populist & elite & 5265 & 51070 & 0.314 & 0.201 & 0.722 \\
environment & 2019 & LLM & climate\_denialist & pro\_environment & 107 & 28163 & 0.011 & 0.005 & 0.626 \\
\hline
\end{tabularx}
\label{tab:validation_full}
\end{table*}

\section{Human annotation protocol}

In our manual annotation process, we are not looking to capture all the richness of expressions with which an individual may declare identification with an ideological or issue stance. On the contrary, we are looking to identify a small but sufficient group of individuals for which the group identification is certain, so that we can use this certainty in assessing their spatial coherence. We are not concerned by the diverse ways by which individuals might communicate that they identify with ideological leaning or stances on issues that speak to the selected CHES dimensions. 
For instance, our protocol has high chances of identifying a description such as ``I am a proud leftist'' as being on the ``left'', but low chances of identifying a description such as ``I will always follow marxist economic policies'', or cues that rely on niche knowledge or current events.

The following are the criteria given to annotators for the labels used in validations.

\paragraph{Left and Right.}
Criterion: only explicit identification with the ``left'' of the ``right''. Examples: ``Teacher, always on the left'' (left); ``Neither left nor right'' (neither); ``I hate leftists'' (neither, this is not a right-winger necessarily);  ``Husband, parent. Conservative'' (neither); ``I vote for NUPES'' (neither, we are interested in measuring party identification and left- or right-wing identification independently); ``I’ve always been a marxist'' (this is more complicated: it references belonging to the left via historical markers, and not current political competition. It’s up to the annotator to decide.); ``we should deport all migrants, legal AND illegal'' (neither, this might be expression of stances aligned with left-right, but we are precisely interested at measuring this alignment, rather than relying on it); ``I am on the right side of life'' (it’s up to the annotator, but we’ve seen that on some countries these are coined phrases of right-wing identification). In general, we favor explicit mentions of identification with groups ``left'' and ``right''.

\paragraph{Populists and elites.}
Criterion for populist: the person refers to ``the people'', or ``the elite'' (also simply ``elite''), mentions to ``politicians'' (plural) in a critical way (not specific criticism of a single politician), criticism towards institutions or elites (including references to corruption), or expressed sympathies for strongman rule (e.g. ``If Macron was as brave as Putin we wouldn’t have terrorism in France''). Criterion for elite: is a member of an ``elite group''. This includes several groups theorized in the literature: governing elites, economic elites. We exclude political elites (another group theorized in the literature): council members, majors, senators, deputies, etc. Keep in mind that X is an elite space in itself in many ways, and that we’re interested in differences within our X populations. We aim at labeling as ``elite'' individuals that are part of very selective elites. Examples: high paying jobs, working positions related to finance, tech, international trade, diplomacy. We acknowledge that the definition for the label ``populist'' is conceptually clearer than that for ``elite''.
Rather than relying on a robust and inclusive delimitation, our strategy relies on an exclusive and conservative delimitation.
Users with the label do not capture the totality of users that may display populist views, nor those of elites belonging to elite groups. 
Instead, those that have the label, hold a small probability of being mislabeled and should thus be coherently spatialized along the corresponding dimension.

\paragraph{Eurosceptics and Pro Europeans.}
Criterion for pro European: subscribes to positive views on Europe (includes displaying European symbols, or describing oneself as European, or working on European organization). Criterion for eurosceptic: involves criticism towards the E.U.\ (including negative mentions of Ursula von der Leyen, Brussels or other symbols of the EC and EU institutions).

\paragraph{Favorable and opposed to liberal immigration policies.}
Following the framing of our reference document, the CHES codebook, the criterion for both labels were left unqualified beyond what the name signifies. Annotators were instructed to label profile bios according to whether they displayed views that could be deemed ``favorable''  or ``opposed'' to ``liberal immigration policies''.

\section{LLM annotation protocol}

The number of contradictory LLM labels (value 1 in both opposite categories) was the following: left and right 1733 (4.35\%) ; populist elite 950 (1.64\%) ; eurosceptic and pro\_european 110 (0.51\%) ; liberal\_immigration and restrictive\_immigration 53 (0.61\%)  ; cosmopolitan and nationalist 17 (0.09\%) ; pro\_environment and climate\_denialist 6 (0.02\%)  ; economic\_focus pro\_environment 765 (2.04\%)  ; liberal and conservative 49 (0.34\%). Those labels were discarded and replaced by Nans in the published version of the data. 

We now list all the prompts that were used for the annotation process with the LLM. 

\paragraph{Left.} \emph{You are an expert in European politics. Please classify the following X profile bio as ``Left-leaning'' or ``Not-Left'' according to whether the author of the text (who is from France) is politically Left-leaning or not. The response should be in the form of a single term with the name of the category: ``Left-leaning'' or ``Not-Left'': [TEXT OF THE BIO]}

\paragraph{Right.} \emph{You are an expert in European politics. Please classify the following X profile bio as ``Right-leaning'' or ``Not-Right'' according to whether the author of the text (who is from France) is politically Right-leaning or not. The response should be in the form of a single term with the name of the category: ``Right-leaning'' or ``Not-Right'': [TEXT OF THE BIO] }

\paragraph{Populist.} \emph{You are an expert in European politics. Please classify the following X profile bio as ``Populist'' or ``Not-Populist'' according to whether the author of the text (who is from France) holds populist views or not. Populist views include, among others, believing that society is split between the people and elites, or that political elites are corrupt. The response should be in the form of a single term with the name of the category: ``Populist'' or ``Not-Populist'': [TEXT OF THE BIO] }


\paragraph{Elite.} \emph{You are an expert in European politics. Please classify the following X profile bio as ``Elite'' or ``Not-Elite'' according to whether the author of the text (who is from France) belongs to an elite group, including political or economic elites. The response should be in the form of a single term with the name of the category: ``Elite'' or ``Not-Elite'': [TEXT OF THE BIO] }

\paragraph{Pro European.} \emph{You are an expert in European politics. Please classify the following X profile bio as ``Pro-European'' or ``Not-Pro-European'' according to whether the author of the text (who is from France) holds positive views of the European Union or not. The response should be in the form of a single term with the name of the category: ``Pro-European'' or ``Not-Pro-European'': [TEXT OF THE BIO]}

\paragraph{Eurosceptic.} 
\emph{
You are an expert in European politics. Please classify the following X profile bio as ``Eurosceptic'' or ``Not-Eurosceptic'' according to whether the author of the text (who is from France) holds negative views of the European Union or not. The response should be in the form of a single term with the name of the category: ``Eurosceptic'' or ``Not-Eurosceptic'': [TEXT OF THE BIO]}

\paragraph{Conservative.} 
\emph{
You are an expert in European politics. Please classify the following X profile bio as ``Conservative'' or ``Not-Conservative'' according to whether the author of the text (who is from France) holds conservative views or beliefs, including but not limited to negative views on abortion, gender equality, and same-sex marriage. The response should be in the form of a single term with the name of the category: ``Conservative'' or ``Not-Conservative'': [TEXT OF THE BIO]}

\paragraph{Liberal.} 
\emph{
You are an expert in European politics. Please classify the following X profile bio as ``Liberal'' or ``Not-Liberal'' according to whether the author of the text (who is from France) holds liberal views or beliefs, including but not limited to positive views on abortion, gender equality, and same-sex marriage. The response should be in the form of a single term with the name of the category: ``Liberal'' or ``Not-Liberal'': [TEXT OF THE BIO]}

\paragraph{Liberal immigration.} 
\emph{
You are an expert in European politics. Please classify the following X profile bio as ``Pro-Immigration'' or ``Not-Pro-Immigration'' according to whether the author of the text (who is from France) holds liberal or positive views on immigration. The response should be in the form of a single term with the name of the category: ``Pro-Immigration'' or ``Not-Pro-Immigration'': [TEXT OF THE BIO]}

\paragraph{Restrictive immigration.} 
\emph{
You are an expert in European politics. Please classify the following X profile bio as ``Anti-Immigration'' or ``Not-Anti-Immigration'' according to whether the author of the text (who is from France) holds conservative or negative views on immigration. The response should be in the form of a single term with the name of the category: ``Anti-Immigration'' or ``Not-Anti-Immigration'': [TEXT OF THE BIO]}

\paragraph{Pro environment.} 
\emph{
You are an expert in European politics. Please classify the following X profile bio as ``Pro-Environment'' or ``Not-Pro-Environment'' according to whether the author of the text (who is from France) supports environmental protection, including but not limited to fighting against climate change, supporting economic de-growth, showing interest in the rights of animals, or adopting dietary restrictions such being vegan or vegetarian. The response should be in the form of a single term with the name of the category: ``Pro-Environment'' or ``Not-Pro-Environment'': [TEXT OF THE BIO]}

\paragraph{Climate denialist.} 
\emph{
You are an expert in European politics. Please classify the following X profile bio as ``Climate-Denialist'' or ``Not-Climate-Denialist'' according to whether the author of the text (who is from France) denies or is skeptical about climate change, climate warning, or their human cause. The response should be in the form of a single term with the name of the category: ``Climate-Denialist'' or ``Not-Climate-Denialist'': [TEXT OF THE BIO]}

\paragraph{Economic focus.} 
\emph{
You are an expert in European politics. Please classify the following X profile bio as ``Economic-Focus'' or ``Not-Economic-Focus'' according to whether the author of the text (who is from France) is interested in the economy, including but not limited to economic growth. The response should be in the form of a single term with the name of the category: ``Economic-Focus'' or ``Not-Economic-Focus'': [TEXT OF THE BIO]}

\paragraph{Nationalist.} 
\emph{
You are an expert in European politics. Please classify the following X profile bio as ``Nationalist'' or ``Not-Nationalist'' according to whether the author of the text (who is from France) is a nationalist or a patriot, including but not limited to opposing multiculturalism, international organizations, or ethnic minority communities. The response should be in the form of a single term with the name of the category: ``Nationalist'' or ``Not-Nationalist'': [TEXT OF THE BIO]}

\paragraph{Cosmopolitan.} 
\emph{
You are an expert in European politics. Please classify the following X profile bio as ``Cosmopolitan'' or ``Not-Cosmopolitan'' according to whether the author of the text (who is from France) is cosmopolitan, including but not limited to supporting multiculturalism, international organizations, international integration, and multilateralism. The response should be in the form of a single term with the name of the category: ``Cosmopolitan'' or ``Not-Cosmopolitan'': [TEXT OF THE BIO]}
\end{appendices}

\end{multicols}

\end{document}